\documentclass[acmsmall,nonacm]{acmart}
\usepackage{graphicx}
\usepackage{soul}

\usepackage{array}
\usepackage{booktabs}
\usepackage{multirow}
\usepackage{mathtools}

\usepackage{url}
\usepackage{verbatim}
\begin{document}
%
\title[Specification Architectural Viewpoint for Benefit-Cost-Risk-Aware Decision-Making in Self-Adaptive Systems]{Specification Architectural Viewpoint for Benefit-Cost-Risk- Aware Decision-Making in Self-Adaptive Systems}

\author{Danny Weyns}
\affiliation{
	\institution{Katholieke Universiteit Leuven}
	\country{Belgium}
}
\affiliation{
	\institution{Linnaeus University}
	\country{Sweden}
}
\email{danny.weyns@kuleuven.be}

\author{Paris Avgeriou}
\affiliation{
	\institution{University of Groningen}
	\country{The Netherlands}
}

\author{Radu Calinescu}
\affiliation{
	\institution{University of York}
	\country{UK}
}

\author{Sara M Hezavehi}
\affiliation{
	\institution{University of Groningen}
	\country{The Netherlands}
}
\affiliation{
	\institution{Linnaeus University}
	\country{Sweden}
}

\author{Raffaela Mirandola}
\affiliation{
	\institution{Politecnico di Milano}
	\country{Italy}
}

\author{Diego Perez-Palacin}
\affiliation{
	\institution{Linnaeus University}
	\country{Sweden}
}

%
\begin{abstract}
\textbf{Abstract.} Over the past two decades, researchers and engineers have extensively studied the problem of how to enable a software system to deal with uncertain operating conditions. One prominent solution to this problem is \textit{self-adaptation}, which equips a software system with a feedback loop that resolves uncertainties during operation and adapts the system to deal with them when necessary. Most self-adaptation approaches developed so far use decision-making mechanisms that focus on achieving a set of goals, i.e., that select for execution the adaptation option with the best \textit{estimated benefit}. A few approaches have also considered the \textit{estimated (one-off) cost} of executing the candidate adaptation options. We argue that besides benefit and cost, decision-making in self-adaptive systems should also consider the \textit{estimated risk} the system or its users would be exposed to if an adaptation option were selected for execution. Balancing all three factors when evaluating the options for adaptation when mitigating uncertainty is essential, not only for satisfying the concerns of the stakeholders, but also to ensure safety and public acceptance of self-adaptive systems. In this paper, we present an ISO/IEC/IEEE 42010 compatible architectural viewpoint that considers the estimated benefit, cost, and risk as core factors of each adaptation option considered in self-adaptation. The viewpoint aims to support software architects responsible for designing robust decision-making mechanisms for self-adaptive systems.
\end{abstract}

\maketitle 
\section{Introduction}

Modern software systems are expected to deal with changing operating conditions. Examples are dynamic workloads, fluctuating availability of services, and changes in the environment of the system. For  designers, these changes create uncertainties that may be difficult to anticipate before deployment~\cite{Ramirez2012,Esfahani2013,icpe14,Mahdavi2017,Calinescu2020}. Yet, without mitigation, such uncertainties may jeopardise the concerns of the stakeholders of the system. One approach to mitigate uncertainties is self-adaptation~\cite{cheng2009software,Weyns2019}. Self-adaptation extends a system with a feedback loop that tracks the system and its environment to resolve the uncertainties and adapt the system to deal with the changing conditions, or gracefully degraded if necessary. 
Over the past two decades, researchers and engineers have extensively studied the problem of how to realise self-adaptation using feedback loops~\cite{Weyns2019}. 

A common approach is to equip the feedback loop with a decision-making mechanism that evaluates the relevant options for adaptation (i.e., the possible configurations of the system that are considered to adapt the system) and selects the option with the best expected outcome in terms of achieving the system goals. We refer to this as the \textit{estimated benefit} that can be achieved by self-adaptation when a particular adaptation option is selected for execution. 
Consider for instance 
an e-health system that remotely monitors vital parameters of elderly people and takes action when needed; for instance, change drugs, visit and support the elderly, or send an emergency team. To perform the analysis of vital parameters or dispatch health carers when needed, the  system uses third-party services. The load of the system (i.e., service requests) may dynamically change in ways that may be difficult to predict; for instance, the health conditions of elderly may be affected by changing weather. To deal with such changes, the system may dynamically adapt the allocation of service requests to third-party services such that the response time of the service remains below a given threshold, while the reliability of the system is maximised. To that end, the system will estimate the benefit in terms of the values of the required quality attributes (i.e., response time and reliability in our example) for different compositions of services provided by the service providers, and then adapts the system such that it selects the best composition of services~\cite{7194661}. 

A few approaches also take into account the  \textit{estimated cost} for adapting the system when evaluating the options for adaptation, see for instance~\cite{5541703,VanDerDonckt2018}. In this context, we use `cost' to refer to the one-off cost to perform an adaptation of the running system.\footnote{Note that the one-off cost for adaptation contrasts with other aspects such as the financial cost for the owner to run the application, or the price that may need to be payed by users when using the service. If deemed relevant for the stakeholders, these aspects need to be considered when determining the expected benefits of adapting the system.} As an example, switching services in the e-health system may require the system to test the reliability of newly selected services before using them. Such tests require extra resources and time. The level of testing that is required may depend on the service-level agreement of the service providers of the services under test. When assessing a composition of services provided by third-parties, the system will calculate the estimated cost for testing the different newly integrated services of the  composition. Hence, when making adaptation decisions, the system should not only consider the estimated benefit of the each of the adaptation options, but should also take into account the estimated cost of the different service compositions that can be selected. 

Besides estimated benefit and cost, we argue that adaptation decisions should also take into account the \textit{estimated risk} the system would be exposed to if an adaption option were selected and used to adapt the running system. Considering that risk is particularly important in domains where decision-making may affect the safety and/or privacy of users, have an impact on the environment, or jeopardise ethical or legal concerns~\cite{Fischhoff2015}. 
With the advancing digitisation of society and industry, these aspects apply to a wide variety of computing systems that are becoming vital for our society. 
Compared to other areas where risks are taken into account in decision-making, see for instance~\cite{62930,Alali2012,6606612,holstein2018ethical}, the decision-making of self-adaptive systems is lagging behind. Yet, compared to other areas, where risk analysis is done by humans supported by tools, risk analysis in self-adaptive systems needs to be done automatically by the self-adaptive system itself within the time window that is available to take adaptation decisions. Clearly, this calls for solid preparation during system design, such that adaptation decisions that take risk into account can be made efficiently by the system during operation.   
%
This paper aims at making a step forward towards dealing with risk in self-adaptive systems.   
We illustrate the importance of considering estimated risk in the e-health system, which is an example of a  valuable computing system for society. Selecting a new composition of services from different service providers may affect the privacy of the elderly people that use the services. In particular, different service providers may apply different rules regarding the collection and treatment of data from the e-health system. Hence, the selection of a particular set of services 
implies a particular data privacy risk for the system users. When selecting services, the system should estimate this risk, e.g., based on the trustworthiness of the service providers.
Hence, decision-making in self-adaptation should in general take into account the estimated benefit, cost, and risk of the different options considered for adaptation. 

This paper presents an architectural viewpoint for 
decision-making in self-adaptive systems that takes into account estimated benefit, cost, and risk as first-class citizens. 
The primary users that will benefit from the viewpoint are the software architects of self-adaptive systems. The viewpoint is structured using the template recommended by the ISO/IEC/IEEE 42010 standard~\cite{ISO2011}. 
This standard defines an architecture viewpoint as ``a work product establishing the conventions for the construction, interpretation and use of architecture views to frame specific system concerns.'' An architectural viewpoint gives an architect the means to express a coherent set of concerns, the stakeholders interested in these concerns, and \emph{model kinds} (i.e., meta-models) that frame the concerns, each defining notations, modelling templates, analytical methods and possibly other operations useful on models of the model kind. A viewpoint can be instantiated for a domain at hand, resulting in a view that comprises architecture models that address the concerns framed by its governing viewpoint. As such, the viewpoint presented in this document establishes the conventions for defining and using architecture views to deal with the concerns of stakeholders for decision-making in self-adaptation by taking into account the estimated benefit, cost, and risk of adaptation options. 
The viewpoint is grounded in a study of the literature and supported with extensive experiences in engineering self-adaptive systems across a wide range of domains. 

\section{Scope of the Viewpoint}
\label{sec:scope}

Given the wide range of approaches used to realise self-adaptation in general, and decision-making in self-adaptive systems in particular, it is essential that we clarify the definition of the viewpoint before we present its specification.
In this work, we adopt the widely accepted conceptual definitions of architectural viewpoints and models from the IEEE 1471 and ISO/IEC 42010 standards~\cite{4278472}.

The viewpoint is centred on \emph{architecture-based adaptation}~\cite{Oreizy1999,Garlan2004,Kramer2007,Weyns2012}, which is an established approach to engineering self-adaptive systems. Architecture-based adaptation has a dual focus~\cite{weyns2020book}: on the one hand the use of software architecture as an abstraction to \emph{define} a self-adaptive system at design time, and on the other hand the use of architectural models to \emph{reason} about change and make adaptation decisions at runtime. The viewpoint presented in this paper focusses on the second aspect. 
Architecture-based adaptation makes a distinction between domain concerns that are handled by the \textit{managed system}, i.e., what the system should provide to the user in terms of functions or services, and adaptation concerns that are handled by the \textit{managing system}, i.e., how the domain concerns are achieved in terms of benefits, cost, and risk. A reference approach to realise the managing system is by means of a so-called MAPE feedback loop~\cite{Kephart2003,6595487}. MAPE refers to the basic functions realised by the feedback loop: Monitor the system and its environment, Analyse the situation and the options for adaptation, Plan the adaptation of the managed system for the best option, and Execute the actions of the plan to adapt the managed system. The MAPE functions share a repository with Knowledge that typically comprises different types of runtime models~\cite{Weyns2012} (MAPE is therefore sometimes also referred to as MAPE-K). It is important to note that MAPE provides a reference model that describes the essential functions of a managing system and the interactions between them. A concrete architecture maps the functions to corresponding components, which can be a one-to-one mapping or any other mapping, such as a mapping of the analysis and planning functions to one integrated decision-making component. 

The focus of the viewpoint is on the adaptation concerns, in particular the decision-making process to select a configuration from the possible configurations to adapt the system. 
We refer to the set of possible configurations to select from as the \textit{adaptation options} and refer to the complete set as the \textit{adaptation space}. With \textit{relevant adaptation options} we refer to the adaptation options that are deemed to be relevant and are actually analysed, which can be the complete adaptation space or a subset of it, determined using some heuristic or selection mechanism. 

The viewpoint is concerned with uncertainties related to \textit{anticipated change}, i.e., the architect has knowledge of the types of changes that may occur, but not when these changes occur and in what way they may occur~\cite{10.1145/3487921} (for instance the frequency of changes or their intensity). Decision-making in the viewpoint is defined based on abstract functions associated with the estimated benefit, cost, and risk of adaptation options. This allows to support different types of decision-making mechanisms, 
for instance based on rules, 
utilities, or softgoals. 

The approaches used for estimating benefit, cost, and risk build on the Cost Benefit Analysis Method (CBAM)~\cite{CBAMWebsite} and the IEC 31010:2019 standard on
risk management and risk assessment techniques~\cite{RiskWebsite}.
CBAM is an established method for analysing the benefits and costs of architectural designs of software-intensive systems. CBAM takes into account the uncertainty factors regarding benefits and costs, providing a basis for informed decision-making about architectural design or upgrade. In contrast to CBAM, we require an automated approach that makes adaptation decisions for the system at runtime to deal with uncertainties. 
On the other hand, the IEC 31010:2019 standard on risk management and risk assessment provides guidance on the selection and application of various techniques that take into account risk in the decision-making process when mitigating uncertainty. Whereas the standard focuses on techniques that are used to aid decision-making under uncertainty in general, in this viewpoint we require techniques that can be applied automatically by a system at runtime to make adaptation decisions under uncertainty.

\section{Running Example}
\label{sec:example}

\begin{figure}[h!]
	\centering
	\includegraphics[width=0.75\textwidth]{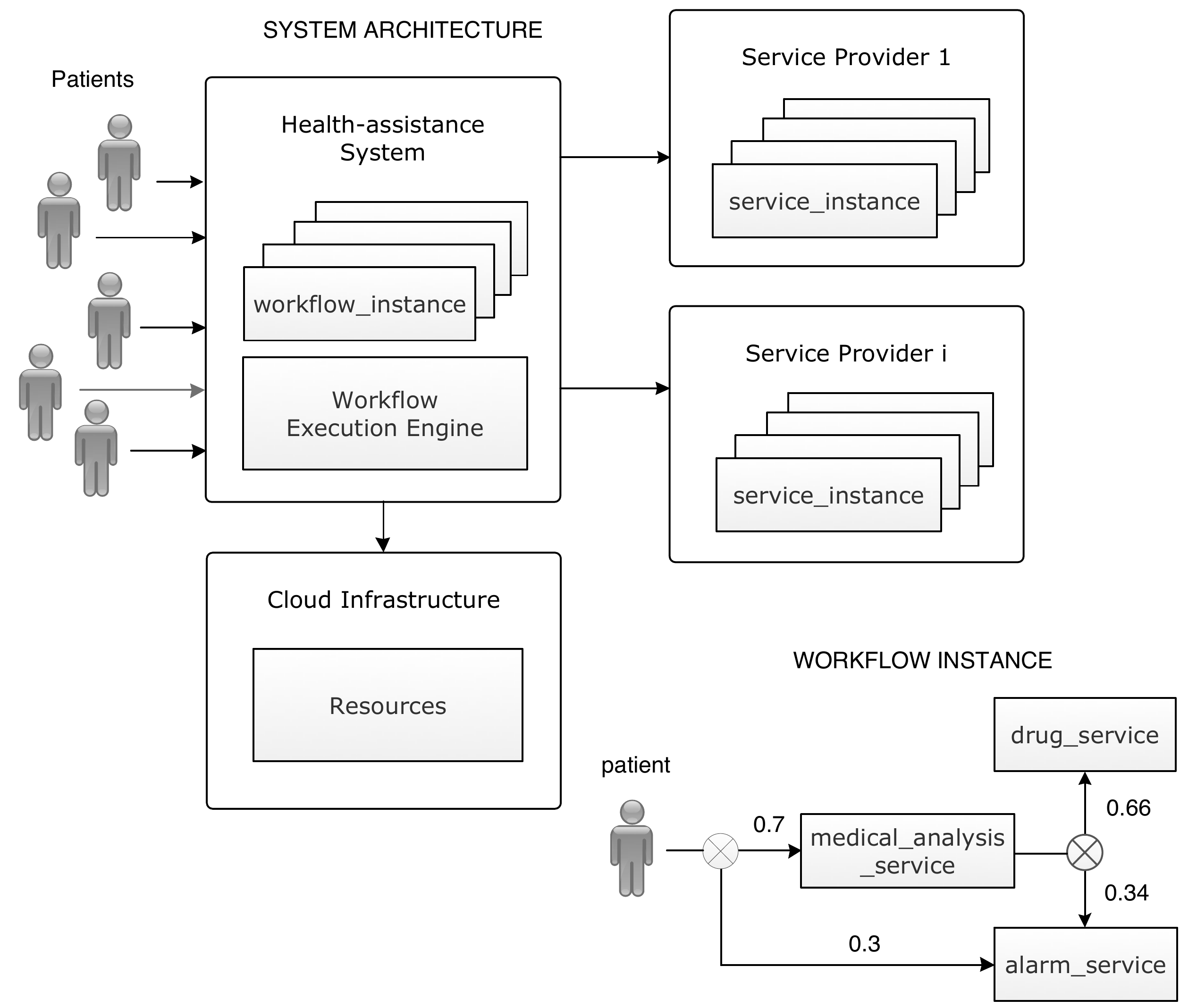}
	\caption{Service-based system: system architecture and workflow instance}
	\label{fig:example}
\end{figure}

We illustrate the different parts of the architectural viewpoint using a classic example from the literature~\cite{7194661}: the e-health system that we already used in the introduction. Consider a simple service-based system as shown in Figure~\ref{fig:example}. This system offers a remote health-assistance service to patients, and relies on data collected via wearable devices. The health-assistance service is realised by a set of specific services that are executed in a workflow as shown in the figure. The core of the application exploits resources of a cloud infrastructure. Each request of a patient instantiates a new instance of the health-assistance service workflow. This workflow then interacts with the services, following the invocation pattern defined by the workflow. A \textit{Medical Service} receives messages with values of vital parameters from the patient's health device. The service analyses the data, and depending on the analysis results nothing needs to be done or action is required. In the latter case, the medical service instructs a \textit{Drug Service} to notify a local pharmacy to deliver new medication to the patient or change the dose of medication, or it instructs an \textit{Alarm Service} in case of an emergency to request medical staff to visit the patient. The alarm service can also directly be invoked by a user via a panic button. The numbers associated with arrows in the workflow express probabilities that actions are invoked. These numbers represent uncertainties that may change over time. Each service can be implemented by a number of service providers that offer concrete services according to a service-level agreement that specifies the reliability and accuracy of the service, among other aspects. Some of the properties of services may change at runtime. For example, due to the changing workloads on the provider side or to unexpected network failures, the reliability of a service may deviate from the one specified in its service-level agreement. Each service provider also offers a privacy policy that specifies how patient data is managed. 

At runtime, it is possible to pick any combination of the available services offered by the service providers. The adaptation goals that express the benefits of adaptation are to keep the average failure low, while minimising the resources required to run the e-health service. Switching services in the system may imply a cost associated with the extra resources that are required to test newly selected services before they can be used. 
Given that service providers may use different privacy policies on how patient data is managed, selecting a service from a service provider implies a risk on the confidentially of the data of patients within legal constraints; e.g., kept strictly local, stored with partners, shared with partners, non-specified.  
Finally, medical analysis services perform their analysis based on the measurements of a limited period using bounded computational resources, there is a risk that the diagnoses derived from the analysis results may not not be 100\% accurate. This may indirectly affect the health conditions of the patients. 

\section{Specification of the Viewpoint}
\label{sec:vp}

Based on a thorough analysis of the literature~\cite{icpe14,Mahdavi2017}, a number of surveys of the community~\cite{Calinescu2020,hezavehi2021uncertainty,10.1145/3487921}, and our own experiences with developing self-adaptive systems in a variety of domains, e.g.,~\cite{8008800,WeynsI0M18,VanDerDonckt2018,Weyns2022}, we defined the architecture viewpoint for benefit-cost-risk-aware decision-making in self-adaptive systems. The viewpoint frames the essential concerns of stakeholders with an interest in the runtime decision-making of self-adaptive systems that are subject to uncertainties. In particular, the viewpoint defines a set of model kinds for identifying, designing, and realising a decision-making module for self-adaptation taking into account the estimated benefit, cost, and risk of the relevant adaptation options. 

It is important to put the viewpoint in a broader context of the design of the feedback loop of a self-adaptive system. The focus of the viewpoint is on the decision-making of self-adaptation, which maps to the analysis function and the first part of the planning function (in the MAPE workflow)~\cite{weyns2020book}. The analysis function evaluates the relevant adaptation options and the first part of planning selects the best option based on the estimated values for benefit, cost, and risk of all analysed adaptation options. The planner function then determines the best option to adapt the running system. As such, architects can combine the viewpoint with additional architectural approaches, such as complementary viewpoints or patterns, to deal with other concerns of realising a self-adaptive system, such as monitoring the system and its environment, keeping runtime models up to date, generating adaptation plans, and executing the adaptation actions of a plan.

\subsection{Stakeholders and Concerns}

Table~\ref{tab:vp_overview} shows an overview of the viewpoint with the stakeholders and their concerns.

\begin{table}[!h!]
\renewcommand{\arraystretch}{1.2}
\caption{Viewpoint -- Overview, Stakeholders, and Concerns.}
\label{tab:vp_overview}
\centering
\begin{tabular}{p{13.4cm} }
\toprule
\textbf{Overview:} \\The architecture viewpoint deals with the main stakeholder concerns related to the decision-making of self-adaptive systems that need to handle uncertain but anticipated changes in the environmemt, the system, and its goals. The viewpoint takes into account the estimated benefit, cost, and risk as first-class citizens when selecting adaptation options. The viewpoint offers model kinds that can be instantiated for a problem at hand. The model kinds show the relevant architectural information that is essential to guide a successful design of benefit-cost-risk-aware decision-making modules of  self-adaptive systems.\\
\midrule
\textbf{Stakeholders:}\\
\emph{Architect(s)} who design the  decision-making module of a self-adaptive system. \\
\emph{Owner(s)} who operate the system and offer its service to users.\\
\emph{User(s)} who use (and pay) for the service of the system. \\
\emph{Other(s)} who may be exposed to potential risks implied by the system.
\\
\midrule
\textbf{Concerns:}\\
\emph{C1 - Benefit:} What are the adaptation goals of the self-adaptive system? What are the estimated values of the quality attributes, i.e., benefit attributes, for an adaptation option that correspond with the adaptation goals? How can the adaptation goals be combined to determine the overall estimated benefit of an adaptation option?\\
\emph{C2 - Cost:} What are the different types of cost associated with performing adaptation? How can the cost for each type be quantified? How can the cost for each type be estimated for an adaptation option? How can the cost for different types be combined to determine the overall estimated cost of an adaptation option?\\ 
\emph{C3 - Desirability:} 
How to balance benefit against cost of an adaption option in order to express its desirability? How to compare and rank desirability results? \\ 
\emph{C4 - Risk:} What are the relevant types of risk for the system? How can the risk for each type be quantified? How can the risk for each type be estimated for an adaptation option? How can the estimates of different types of risk be combined to determine the overall estimated risk of an adaptation option?\\
\emph{C5 - Decision-Making:} What options are available for adapting the system from the current configuration? What elements need to be considered when selecting an adaptation option to adapt the current configuration? How to balance desirability of adaptation options with estimated risk when making adaptation decisions?\\
\bottomrule
\end{tabular}
\end{table}

Stakeholders of the viewpoint are \emph{architect(s)}, \emph{owner(s)}, \emph{user(s)}, and \emph{other(s)}. Architects are primarily interested in technical aspects, in particular the design and behaviour of the decision-making module of the self-adaptive system taking into account the estimated benefit, cost, and risk. Owners have a primary interest in the benefits of the self-adaptive system, the costs that may be induced by adaptation, and the risks that may be implied by adaptations of the system.\footnote{In this viewpoint, we use the term \textit{benefit attribute} to refer to different dimensions of estimated benefit; similarly we use the terms \textit{cost attribute} and \textit{risk attribute} to refer to different dimensions of estimated cost and risk respectively.}
Architects and owners have also an interest in the desirability of system configurations; a configuration with a high desirability has a high benefit and a low cost. Users are primarily interested in the benefits provided by the self-adaptive system as well as the risks they may be exposed to. 
Others are those people that may be exposed to potential risks implied by the system, directly or in the environment. 

In summary, the viewpoint addresses the following adaptation concerns of the stakeholders: \emph{benefits} of adaptation (architects, owners, and users), \emph{costs} of adaptation (architects and owners), \emph{desirability} of adaptation (architects, owners), \emph{risks} implied by adaptation (architects, owners, users, and others), and \emph{decision-making} for adaptation (architects).

\vspace{5pt}\noindent 
\textit{\textbf{Example.}} The architects of the health assistance system are the persons who design and oversee the realisation of the decision-making module of the feedback loop that selects adaptation options to adapt the services used by the workflow. The architects' main concern is to ensure that the system makes proper adaptation decisions, i.e., select adaptation options balancing their estimated benefit, cost, and risk. The system owners are the persons who operate the health-assistance service. The main concerns of the system owners are to provide a good service to the users, optimise the resources to operate the system, keeping the cost required for adaptation low, and minimising the exposed risks. The users are the patients that use the service via a wearable device either to analyse vital parameters or to direct alarm a medical team in case of an emergency. Their main concerns are the  reliability of the service and risks in terms of data privacy and health risks of the application. Finally others are people who may be exposed to risks implied by the system directly or indirectly, in particular risks with respect to the data privacy and decisions made by the system. Others can be relatives and friends of patients, care professionals, among other people. 

\subsection{Viewpoint Model Kinds}

The viewpoint comprises five model kinds.  Table~\ref{tab:vp_mk} presents the first three model kinds: \emph{benefit estimation} (MK1), \emph{cost estimation} (MK2), and \emph{benefit-cost analysis} (MK3). Table~\ref{tab:vp_mk2} presents the last two model kinds: \emph{risk estimation} (MK4) and \emph{decision-making} (MK5). 

\begin{figure}[h!]
	\centering
	\includegraphics[width=0.88\textwidth]{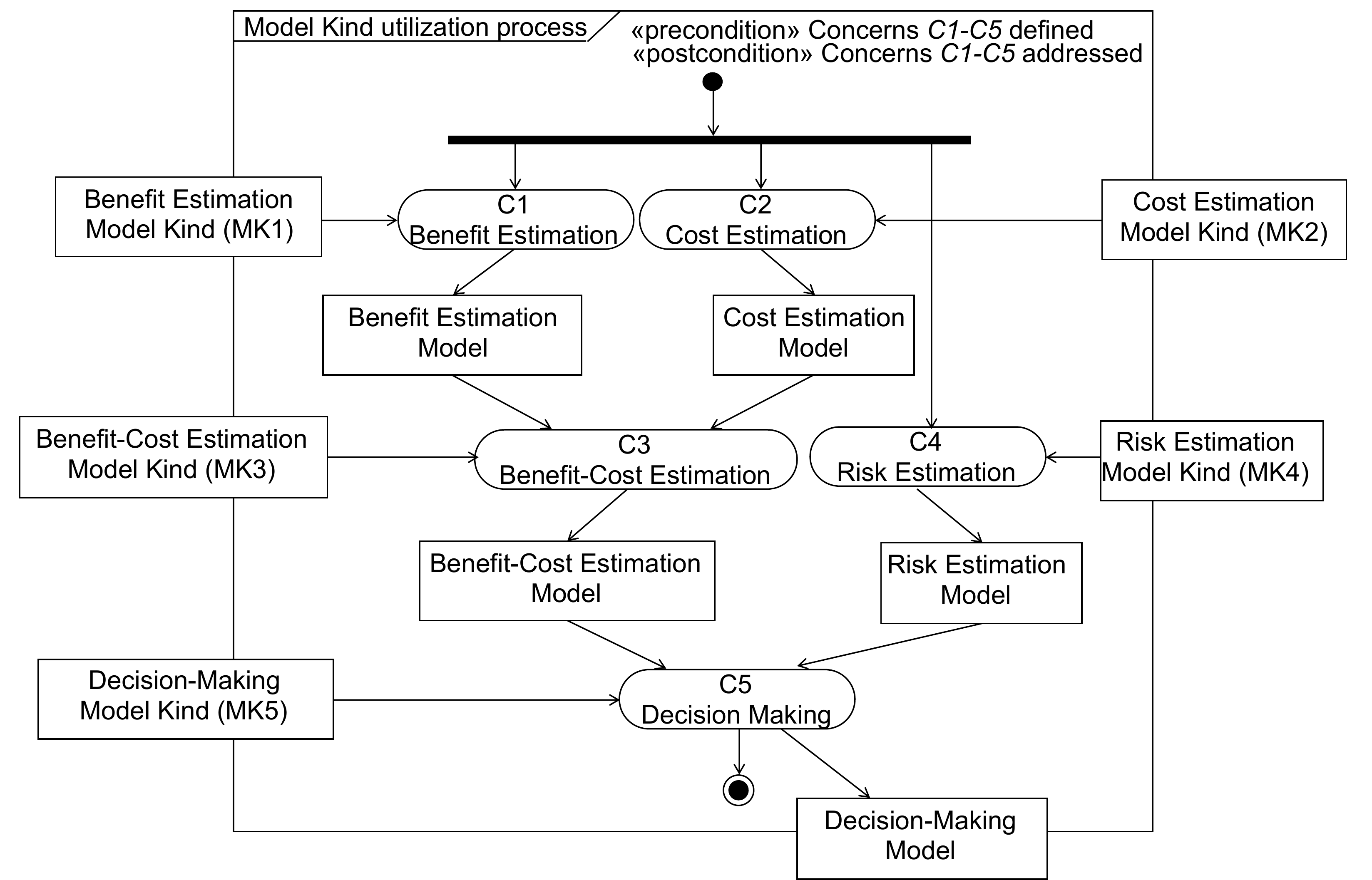}
	\caption{Process to use model kinds.}
	\label{fig:process}
\end{figure}

Figure~\ref{fig:process} gives a high-level overview of the process to use the model kinds. The process starts with the design of the benefit estimation model using the benefit estimation model kind. This model deals with concern C1 (benefit of adaptation options). Next (or in parallel), the  cost estimation model can be designed using cost estimation model kind. This model deals with concern C2 (cost of adaptation options). The next step in the process is the design of the benefit-cost estimation model. This model takes as input the benefit-cost estimation model kind, the benefit estimation model and the cost estimation model. The benefit-cost estimation model deals with concern C3 (desirability of adaptation options). Then (or in parallel), the risk estimation model is designed using the risk estimation model kind. This model deals with concern C4 (risk of adaptation options). In the last step, the decision-making model is designed. This step takes as input the decision-making model kind, the cost-benefit estimation model and the risk estimation model. The decision-making model deals with concern C5 (decision-making, i.e., selecting the best adaptation option). 

\begin{table}[h!t!]
\renewcommand{\arraystretch}{1.2}
\caption{Viewpoint -- Model Kinds MK1, MK2, and MK3}
\label{tab:vp_mk}
\centering
\begin{tabular}{ p{13.4cm} }
\toprule
\textbf{Model Kinds (description):}\\
\emph{MK1 - Benefit Estimation (deals with concern C1):} 
describes per adaptation option how 
each quality attribute is estimated based on the associated adaptation goal and what the overall estimated benefit is based on the combined adaptation goals. \\
\emph{MK2 - Cost Estimation (deals with concern C2):} 
describes per adaptation option how 
each cost attribute is estimated based on the associated cost metric and what the overall estimated cost is based on the combined costs metrics. \\
\emph{MK3 - Benefit-Cost Estimation (deals with concern C3):} 
describes per adaptation option how 
the desirability is determined based a benefit-cost analysis using the estimated benefit and estimated cost of that adaptation option. 
\\
\midrule
\textbf{Model Kinds (meta-models):} 
    \begin{center}
	    \includegraphics[width=0.46\textwidth]{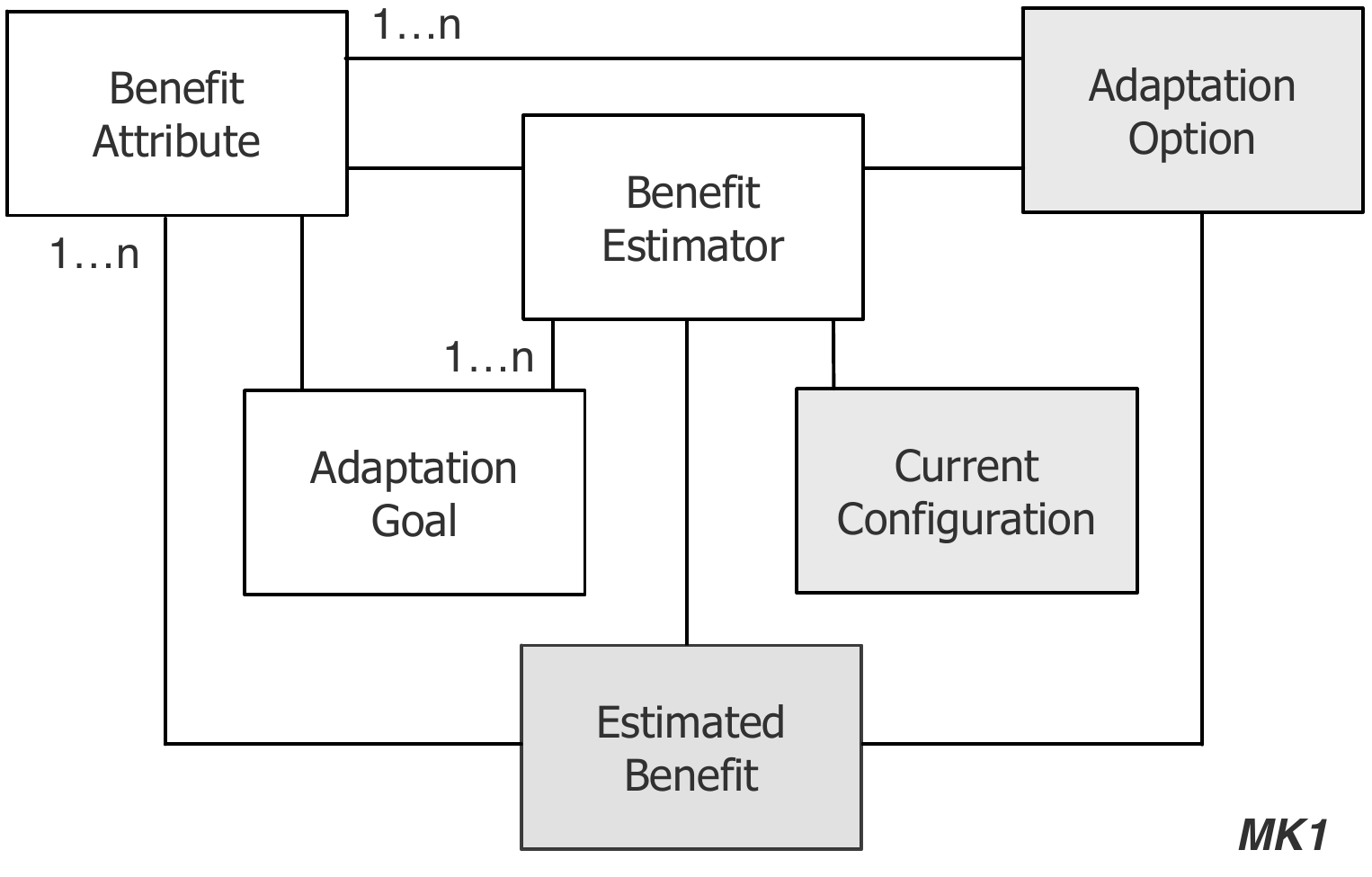}
		\hspace{5pt}
		\includegraphics[width=0.46\textwidth]{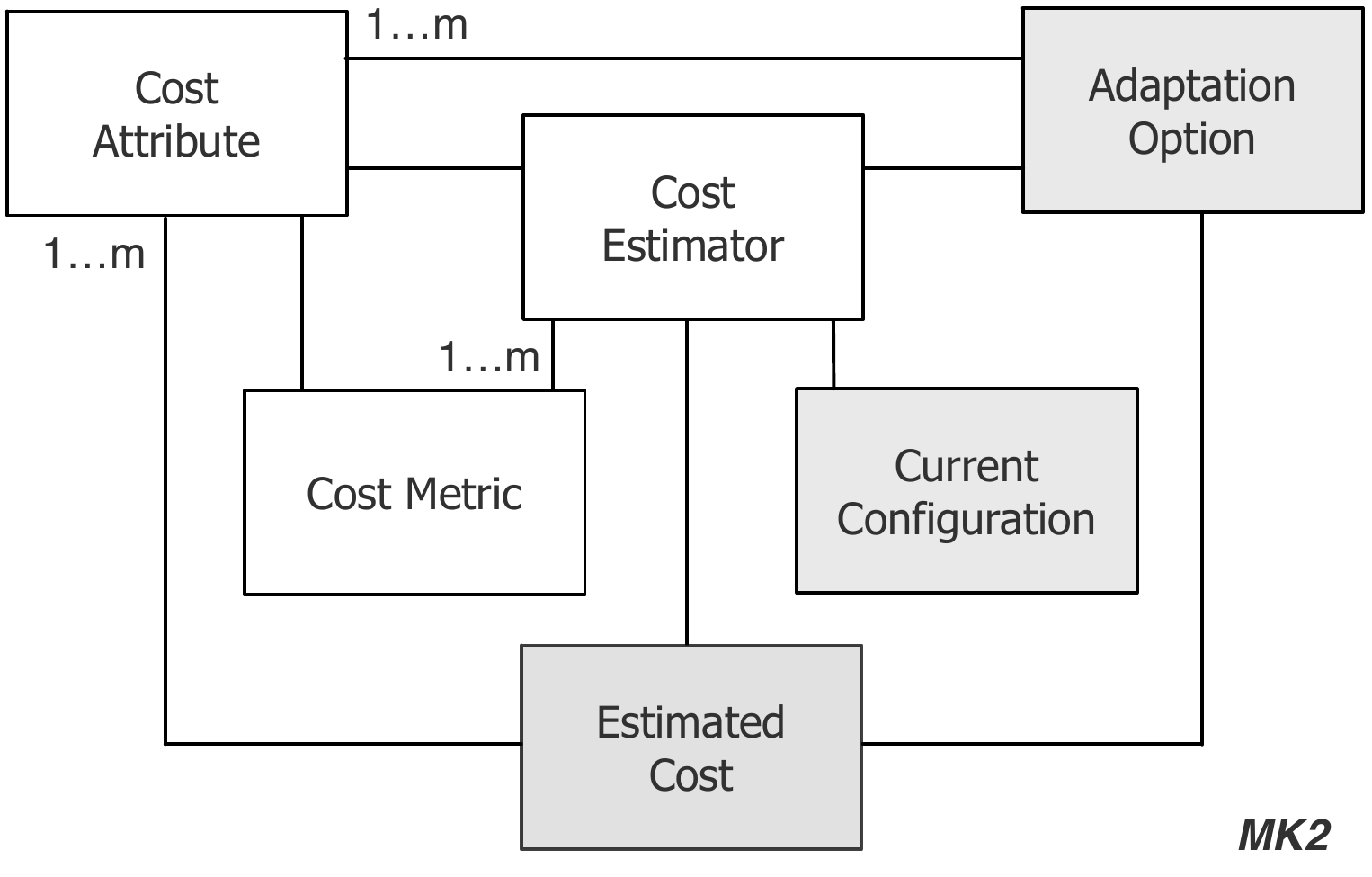}
	\end{center}
	    \begin{center}
		\includegraphics[width=0.46\textwidth]{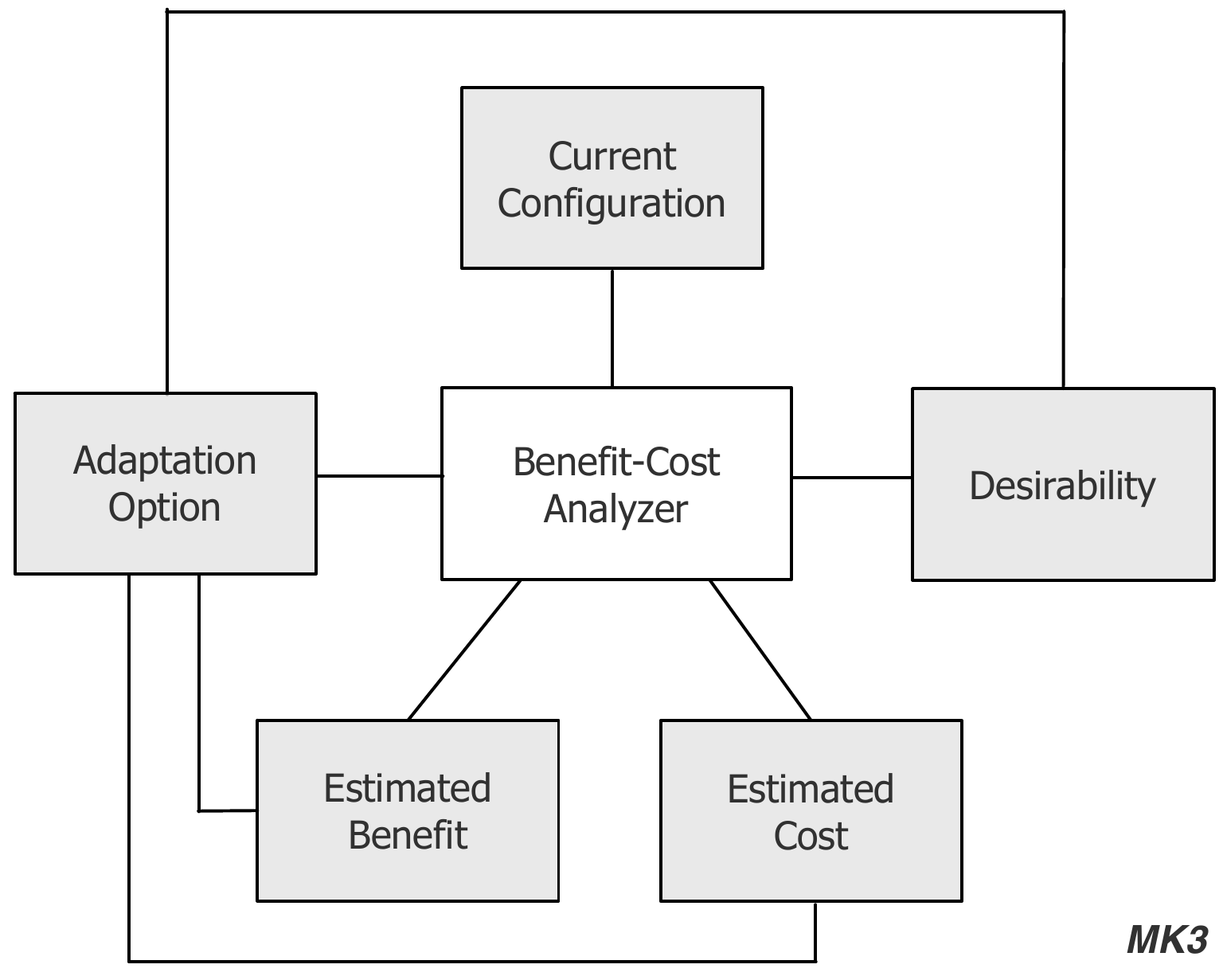}
	\end{center}
	\\ 
\vspace{-12pt} 
 	\emph{Key:} UML (gray boxes represent model elements shared among model kinds)\\
\bottomrule
\end{tabular}
\end{table}

\begin{table}[ht!]
\renewcommand{\arraystretch}{1.2}
\caption{Viewpoint -- Model Kinds MK4 and MK5}
\label{tab:vp_mk2}
\centering
\begin{tabular}{ p{13.4cm} }
\toprule
\textbf{Model Kinds (description):}\\
\emph{MK4 - Risk Estimation (deals with concern C4):} 
describes per adaptation option how 
each risk attribute is estimated based on the associated risk metric and what the overall estimated risk is based on the combined risk metrics. \\
\emph{MK5 - Decision-Making (deals with concern C5):} describes how an adaptation option is selected from the set of adaptation options to adapt the system from its current configuration based on the desirability and estimated risk of the adaptation options.\vspace{5pt}\\
\midrule
\textbf{Model Kinds (meta-models):} 
    \begin{center}
	    \includegraphics[width=0.46\textwidth]{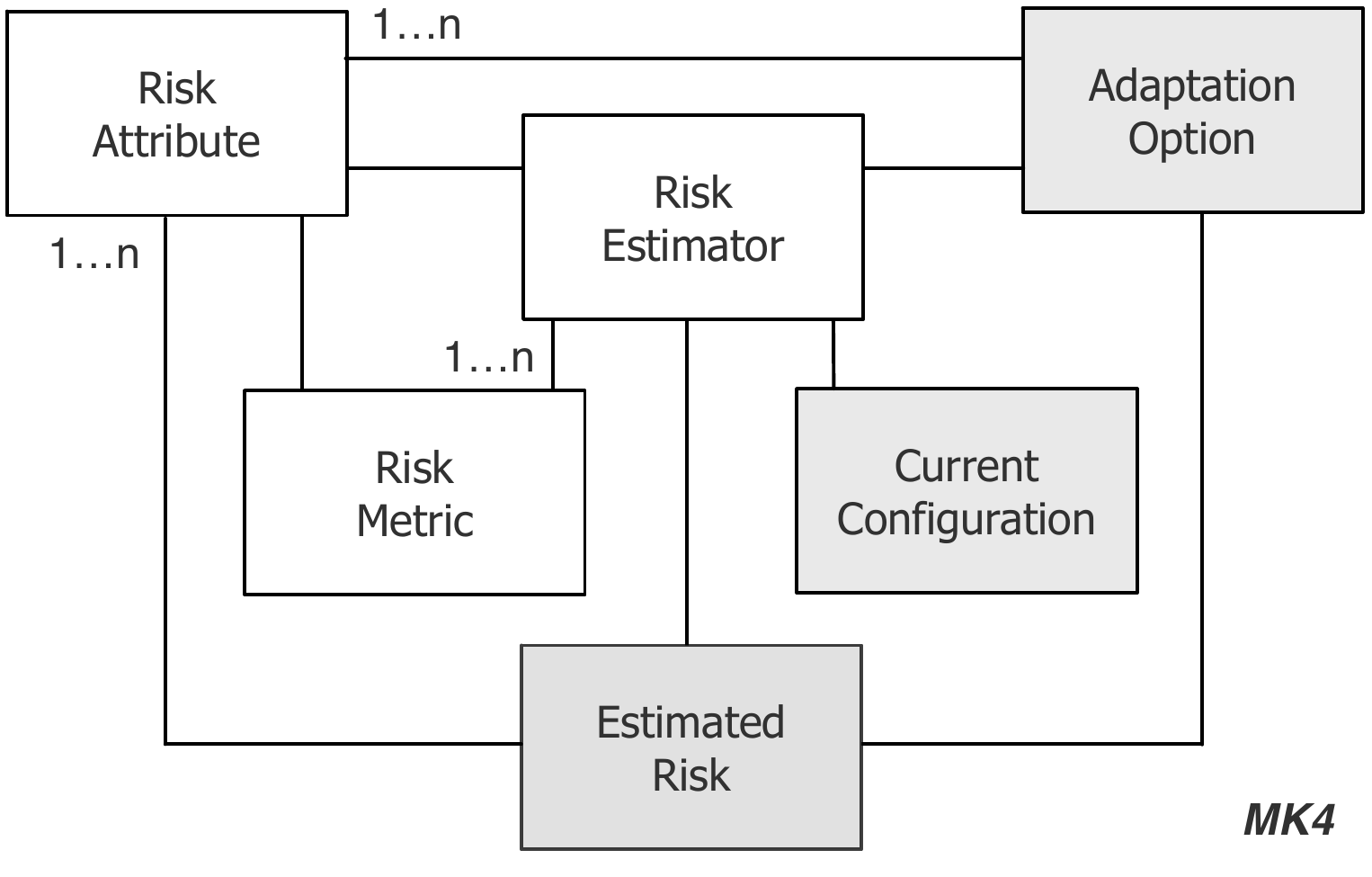}
		\hspace{5pt}
		\includegraphics[width=0.46\textwidth]{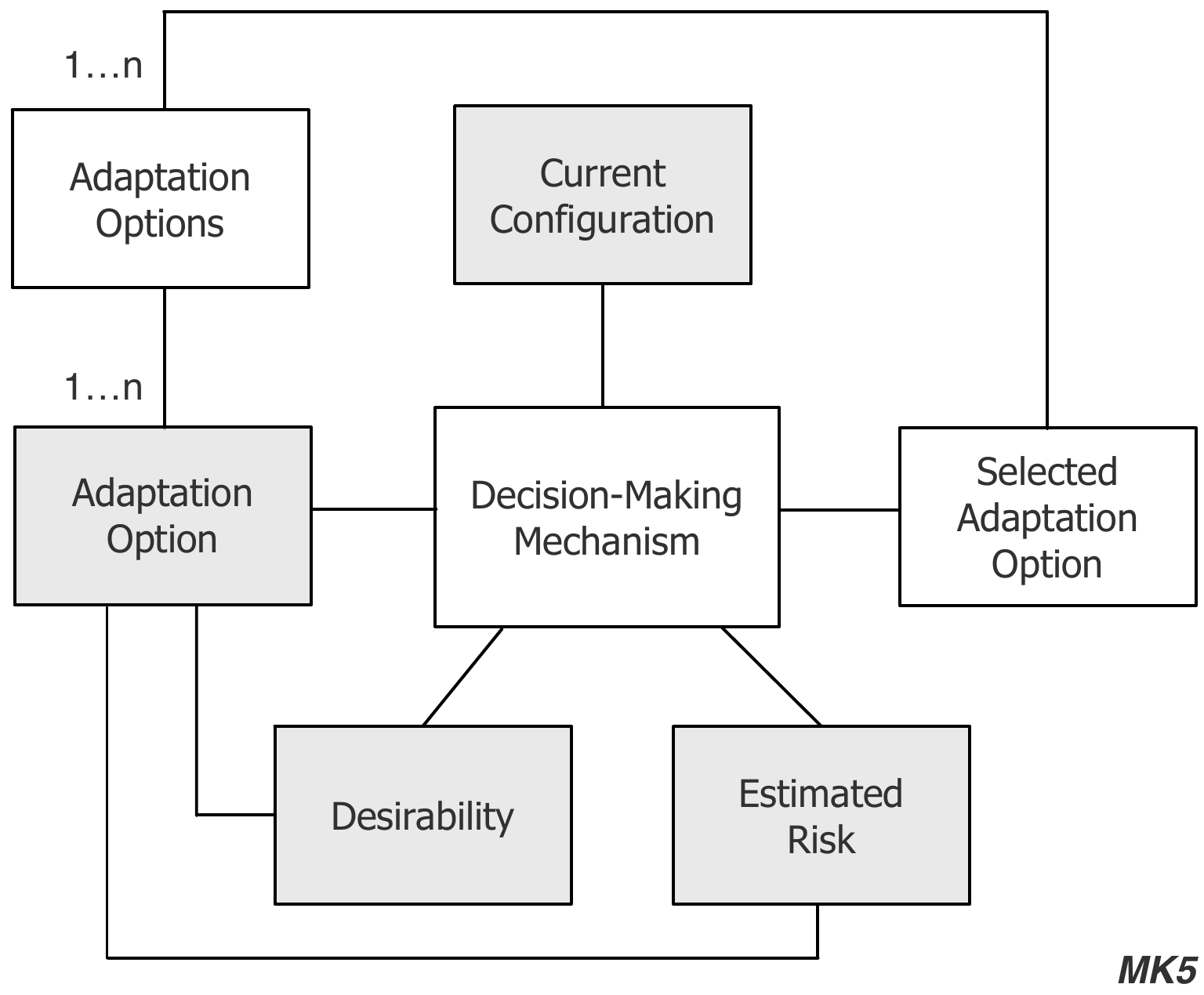}
	\end{center}
	\\ 
\vspace{-12pt} 
 	\emph{Key:} UML (gray boxes represent model elements shared among model kinds)\\
\bottomrule
\end{tabular}
\end{table}

We explain now each model kind in detail and illustrate it with an example of the health-assistance service system. More elaborated descriptions of the core elements used on the model kinds are provided in the Appendix. \\

\noindent\textbf{Benefit Estimation Model Kind (MK1).} This model kind describes how the benefit of each relevant adaption option is estimated (see Table~\ref{tab:vp_mk} top left). The \textit{current configuration} is a representation of the aspects of the managed system and the environment that are relevant to adaptation at that time. These aspects include the current component configuration of the managed system, the settings of relevant system parameters, the values of the quality properties of interest, and the values of uncertainties that are relevant to adaptation. An \textit{adaptation option} is a possible configuration that can be reached from the current configuration by adapting the system. 
An \textit{adaptation goal} represents a requirement that needs to be achieved by the managing system. Adaptation goals usually refer to quality requirements. A \textit{benefit attribute} of an adaptation option represents the estimated value for a system property of the managed system that corresponds with an adaptation goal. Benefit attributes are usually quality properties. With each benefit attribute there is one corresponding adaption goal. 
A \textit{benefit estimator} is a mechanism that enables estimating the benefit attributes of the adaptation options. The \textit{estimated benefit} represents the overall estimated benefit of an adaptation option based on the estimated benefit attributes and combined adaptation goals. 

\begin{figure}[h!]
	\centering
	\includegraphics[height=0.35\textwidth]{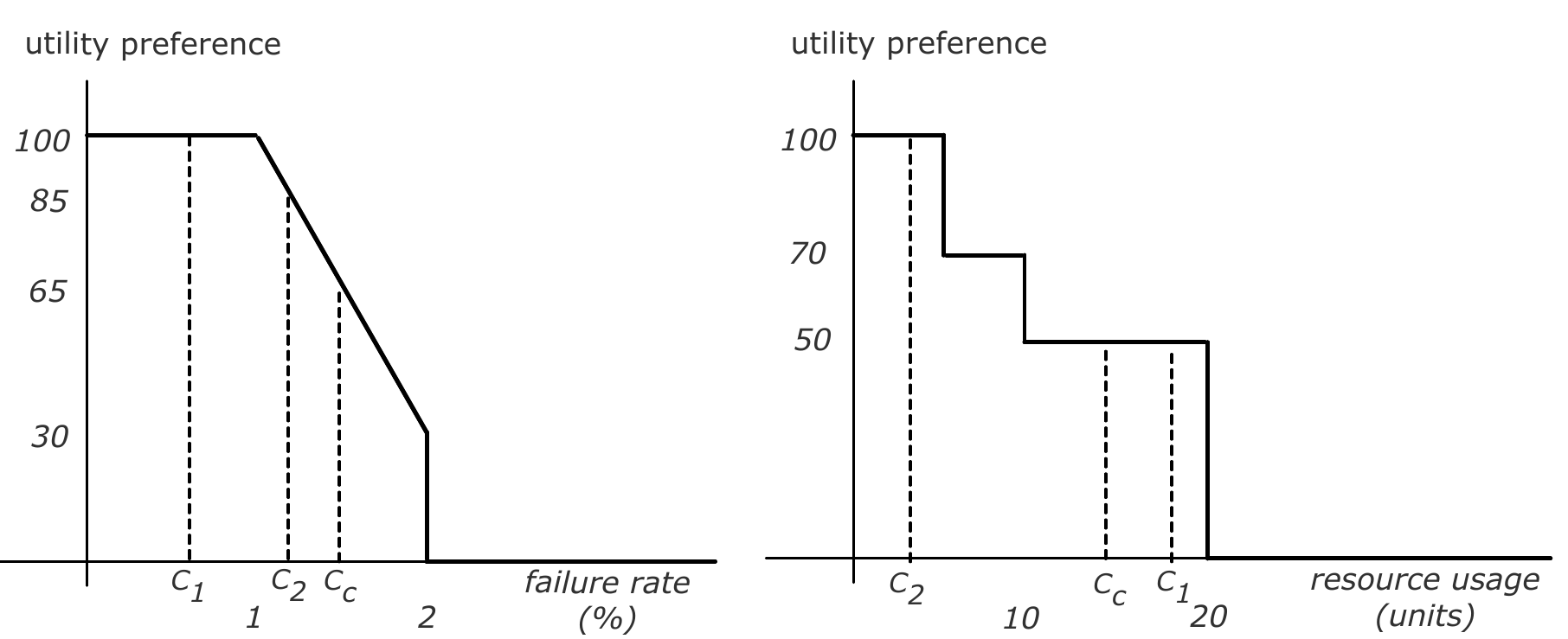}
	\caption{Example utility response curves}
	\label{fig:example_urc}
\end{figure}

\begin{figure}[h!]
	\centering
	\includegraphics[height=0.45\textwidth]{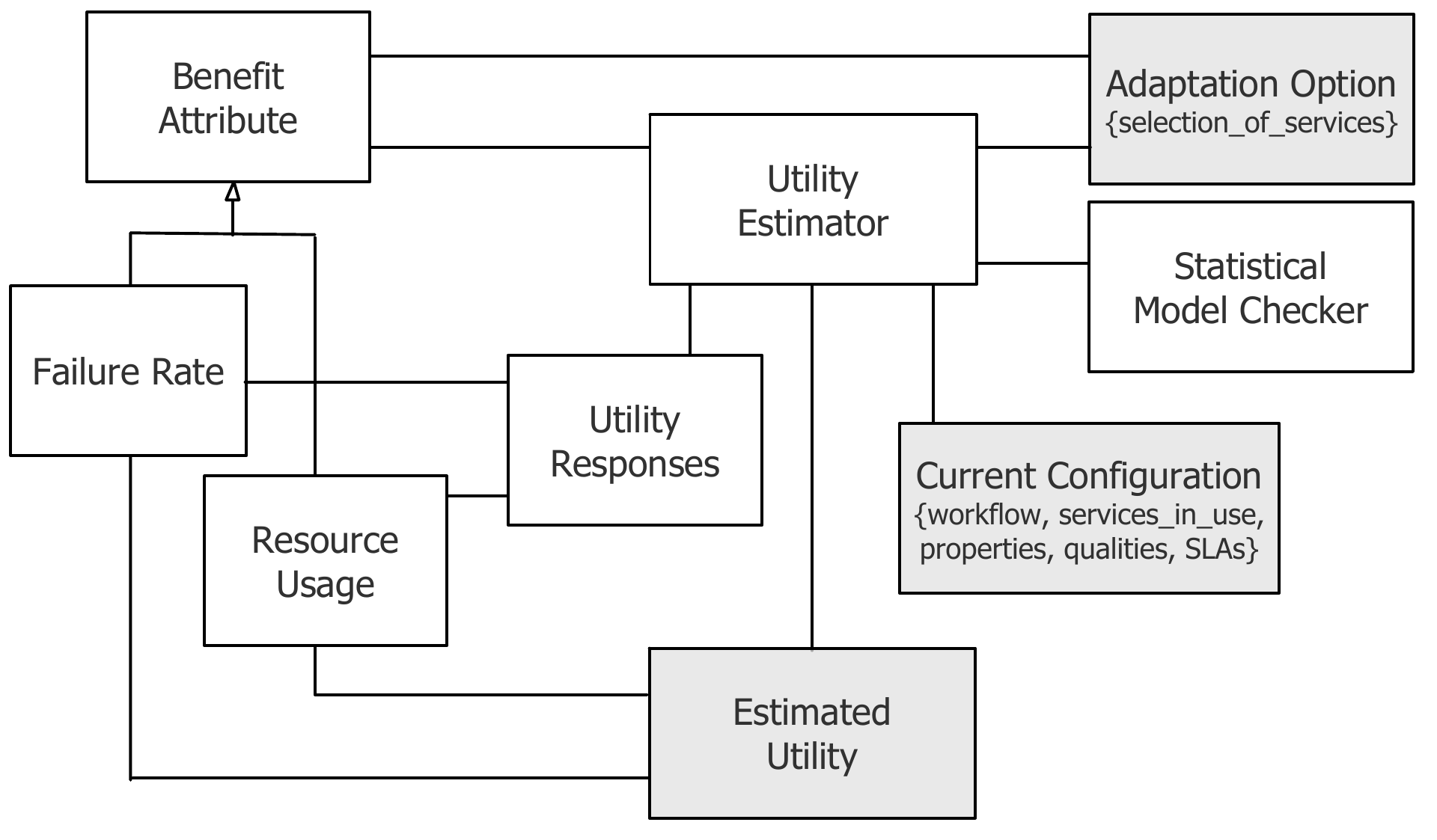}
	\caption{Example instance of Benefit Estimation Model Kind}
	\label{fig:example_MK1}
\end{figure}

\vspace{5pt}\noindent 
\textit{Example.} Figure~\ref{fig:example_MK1} shows an instance of the benefit estimation model kind for the health assistance system. The current configuration consists of the workflow with a set of services currently in use, a set of properties referring to uncertainties including the actual values associated with the different paths exercised in the workflow, the current values of the failure rate and resource usage of the system, and service-level agreements. An adaptation option corresponds to a particular selection of concrete services provided by service providers to be executed by the workflow. The service-based system has two benefit attributes: failure rate and resource usage. The corresponding adaptation goals are represented at utility responses. 
Figure~\ref{fig:example_urc} shows the  utility response curves for both goals. 

As shown in the left part, the utility preference for configurations with failure rates below 1\% is 100\%. For configurations with failure rates between 1 and 2\% the utility preference gradually decreases to 30\%. The utility preference of configurations with failure rates above 2\% is zero. The right part of the figure shows the utility preference for resource usage, defined in a similar way. 
As an example, configuration $C_1$ has a 100\% utility preference for a failure rate of 0.5\% and 50\% for a resource usage of 18 units. Adaptation option $C_2$ has a utility preference of 85\% for a failure rate of 1.3\% and 100\% for a resource usage of 3 units. Current configuration $C_c$ on the other hand has a utility preference of 65\% for a failure rate of 1.5\% and 50\% for a resource usage of 15 units. 
The utility estimator determines the estimated utilities for each relevant adaptation option. To that end, the estimator configures a runtime model of the workflow for each combination of concrete services (adaptation options) together with the actual probabilities that paths are selected. This model is then analysed by a statistical model checker that runs a number of simulations for each adaptation option. The result of the analysis is an estimate for failure rate and resource usage for each adaptation option with a required accuracy and confidence. The estimated benefit is then determined using a utility function that computes the sum of the weighted values of the estimated quality attributes for each adaptation option. Weights express the relative importance of the benefit attributes for the stakeholders. 
In particular, to determine the estimated benefit for an adaptation option we take the sum of the difference between the estimated utility for that adaptation option and the utility of the current configuration for each benefit property taking into account the respective weights. As an example, the estimated benefit of adaptation options can be computed as follows:  

\begin{equation}\label{eq5}
EB_{C_i}= (U_{C_i}^{F} - U_{C_c}^{F}) * W^{F} + (U_{C_i}^{R} - U_{C_c}^{R}) * W^{R}
\end{equation}

with $U_{C_i}^{F}$ the utility of adaptation option $C_i$ for failure rate $F$, $U_{C_c}^{F}$ the utility of the current configuration $C_c$ for the failure rate, and $W^F$ the weight for failure rate (in the example 0.7); the second term with a similar structure refers to resource usage $R$ (with weight 0.3). Applied to adaptation option $C_1$ in Figure~\ref{fig:example_urc} the estimated benefit is:

\begin{equation}\label{eq5}
EB_{C_1}= (100 - 65) * 0.7 + (50 - 50) * 0.3 = 24.5
\end{equation}

Similarly, the benefit of adaptation option $C_2$ in Figure~\ref{fig:example_urc} is:

\begin{equation}\label{eq5}
EB_{C_2}= (85 - 65) * 0.7 + (100 - 50) * 0.3 = 29.0
\end{equation}

In this particular case, adaptation option $C_2$ has a higher estimated benefit as adaptation option $C_1$. Hence, if an adaptation decision would be made based on estimated benefit only, adaptation option $C_2$ would be selected for adaptation. 

\vspace{5pt}
\noindent\textbf{Cost Estimation Model Kind (MK2).} 
This model kind describes how the estimated cost of applying each adaption option is determined (see Table~\ref{tab:vp_mk} top right). A \textit{cost attribute} represents a particular type of cost that is implied by adaptation of the managed system. Different cost attributes can be associated with an adaptation option, such as communication overhead implied by adaptation, extra resources required to perform adaptation, and temporal restrictions in the availability of the system functionality as a consequence of adaptation. \textit{Cost metrics} define measures to quantify cost attributes. A \textit{cost estimator} is a mechanism that enables estimating the cost attributes of the adaptation options. The \textit{estimated cost} represents the overall estimated cost of an adaptation option based on the estimated cost attributes and the combined cost metrics. 

To support architects with identifying cost attributes for a problem at hand, we devised a list of cost attributes with associate cost metrics, see Table~\ref{tab:cost_dim_cost_met}. This table is based on examples extracted from the literature and our own experiences, see for instance~\cite{5541703,VanDerDonckt2018,SEAMS.2019.00023}. 
We distinguish three groups of adaptation costs. The first group refers to resources that are required to realise adaptation, such as bandwidth, processor time, memory, and power. The second group relates to overhead of the system in terms of quality properties that are affected by adaptation, which can be reduced availability, a performance penalty, or costs to deal additional security vulnerabilities implied by adaptation. The third group refers to a monetary price that is implied by the realisation of adaptation. Table~\ref{tab:cost_dim_cost_met} is not meant to be exhaustive and can be easily refined or extended.  

\begin{table}[htbp]
    \centering
        \caption{Cost attributes with cost metrics.}
    \label{tab:cost_dim_cost_met}
    \begin{tabular}{|c|c|c|}
        \hline
        \textbf{Cost type} & 
        \textbf{Cost attribute}     & \textbf{Cost metric} \\ \hline
        Resources & Communication        &   	Required bandwidth  \\ \hline
	    Resources & Computation	      &  Required processing resources \\ \hline
	    Resources & Storage    &  Required memory \\ \hline
	    Resources & Power	&  Required energy \\ \hline	 
	    Overhead & Availability &  Degree of reduced service \\ \hline
        Overhead & Performance & Degree of degraded user experience  \\ \hline
        Overhead & Security &  Cost to manage exposed vulnerability \\ \hline
        Economic & Financial &  Monetary price \\ \hline
    \end{tabular}
    \vspace{5pt}
\end{table}

\begin{figure}[h!]
	\centering
	\includegraphics[height=0.4\textwidth]{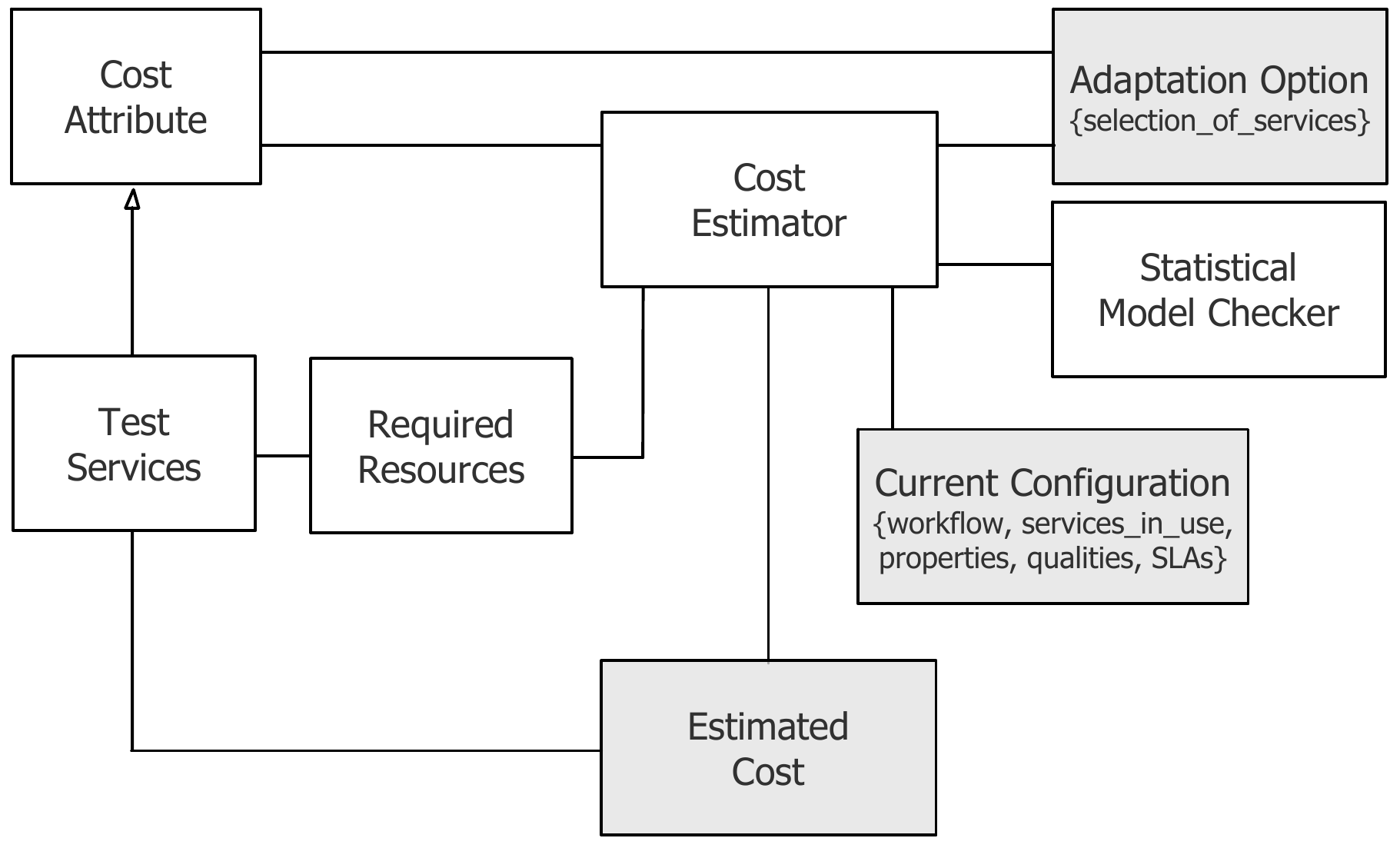}
	\caption{Example instance of Cost Estimation Model Kind}
	\label{fig:example_MK2}
\end{figure}

\vspace{5pt}\noindent 
\textit{Example.} Figure~\ref{fig:example_MK2} shows an instance of the cost estimation model kind
for the health assistance system. We consider one cost attribute: the tests of new services that are considered by the adaptation options~\cite{CalinescuSubmitted}. Testing overhead requires extra resources, which corresponds to computation overhead in Table~\ref{tab:cost_dim_cost_met}. The amount of testing that is needed may depend on trustworthiness of service providers. Table~\ref{tab:cost_dim_cost_met} illustrates a possible cost model. 

\begin{table}[htbp]
    \centering
        \caption{Example cost model for e-health system with required resources for services.}
    \label{tab:cost_dim_cost_met}
    \begin{tabular}{|c|c|c|c|c|}
        \hline
        \textbf{Provider}   & \textbf{SLA} & \textbf{Medical Analysis Service} & \textbf{Drug Service} & \textbf{Alarm Service} \\ \hline
        SP1        &   Silver & 5	& 6 & 2 \\ \hline
        SP2        &   Gold & 3	& 2 & 2 \\
	    \hline
	    SP3        &   Bronze & 8 & 8 & 4 \\
	    \hline
    \end{tabular}
    \vspace{5pt}
\end{table}

Assume that the current configuration comprises the following set of services: 

\begin{equation}\label{eq5}
C_c = \{S_{SP1}^{MAS}, S_{SP3}^{DS}, S_{SP1}^{AS}\}
\end{equation}

$MAS$ is a medical analysis service provided by service provider $SP1$, $DS$ is a drug service provided by $SP3$, and $AS$ is an alarm service provided by $SP1$. Consider now two adaptation options $C_1$ and $C_2$ composed as follows: 

\begin{equation}\label{eq5}
C_1 = \{S_{SP1}^{MAS}, S_{SP2}^{DS}, S_{SP2}^{AS}\};\mbox{\ \ } C_2 = \{S_{SP1}^{MAS}, S_{SP1}^{DS}, S_{SP1}^{AS}\}
\end{equation}

The estimated cost to test the adaptation option of configuration $C_1$ is: \vspace{-5pt}

\begin{equation}\label{eq6}
EC_{C_1} = cost(C_c, C_1) = 
cost_{adapt}(S_{SP3}^{DS},S_{SP2}^{DS}) + cost_{adapt}(S_{SP1}^{AS},S_{SP2}^{AS}) = 2 + 2 = 4 
\end{equation}

The cost only applies to $DS$ and $AS$, i.e., the services that need to be tested before they can be adapted. The cost for testing the different services can be found in Table\,\ref{tab:cost_dim_cost_met}. Similarly, t]he estimated cost to test the adaptation option of configuration $C_2$ is: 

\begin{equation}\label{eq7}
EC_{C_2} = cost(C_c, C_2) = cost_{adapt}(S_{SP3}^{DS}, S_{SP1}^{DS}) = 6 
\end{equation}

Despite the fact that adaptation option $C_1$ requires testing two new services and $C_2$ requires to test only one new services, the estimated cost of $C_1$ is smaller as the estimated cost of $C_2$. The reason is that the new services of $C_1$ are provided by a service provider with a gold service level agreement, requiring less extensive testing of services. In sum, if an adaptation decision would be made based on estimated cost only, adaptation option $C_1$ would be selected for adaptation.


\vspace{5pt}
\noindent\textbf{Benefit-Cost Analysis Model Kind (MK3).} This model kind describes how the \textit{estimated desirability} of each adaptation option is determined (see Table~\ref{tab:vp_mk} bottom). Estimated desirability $ED_{C_i}$ expresses the degree that stakeholders prefer the selection of an adaptation option $C_i$ over other options by comparing total estimated benefit with total estimated cost. The desirability of adaptation options increases with higher estimated benefit and lower estimated cost. The \textit{benefit-cost analyser} computes the desirability of an adaptation option. The computation of desirability requires that the estimated cost and benefit are expressed in a common metric and are scaled to be comparable. Different approaches exist to determine desirability. One common approach is so called ``Value-For-Cost'' that defines desirability as the ratio of estimated overall benefit to estimated overall cost (both scaled). Another approach subtracts the total estimated cost from the total estimated benefit to determine the desirability of an adaptation option. More advanced approaches may include regression and forecasting techniques to determine desirability. 

\begin{figure}[h!]
	\centering
	\includegraphics[width=0.6\textwidth]{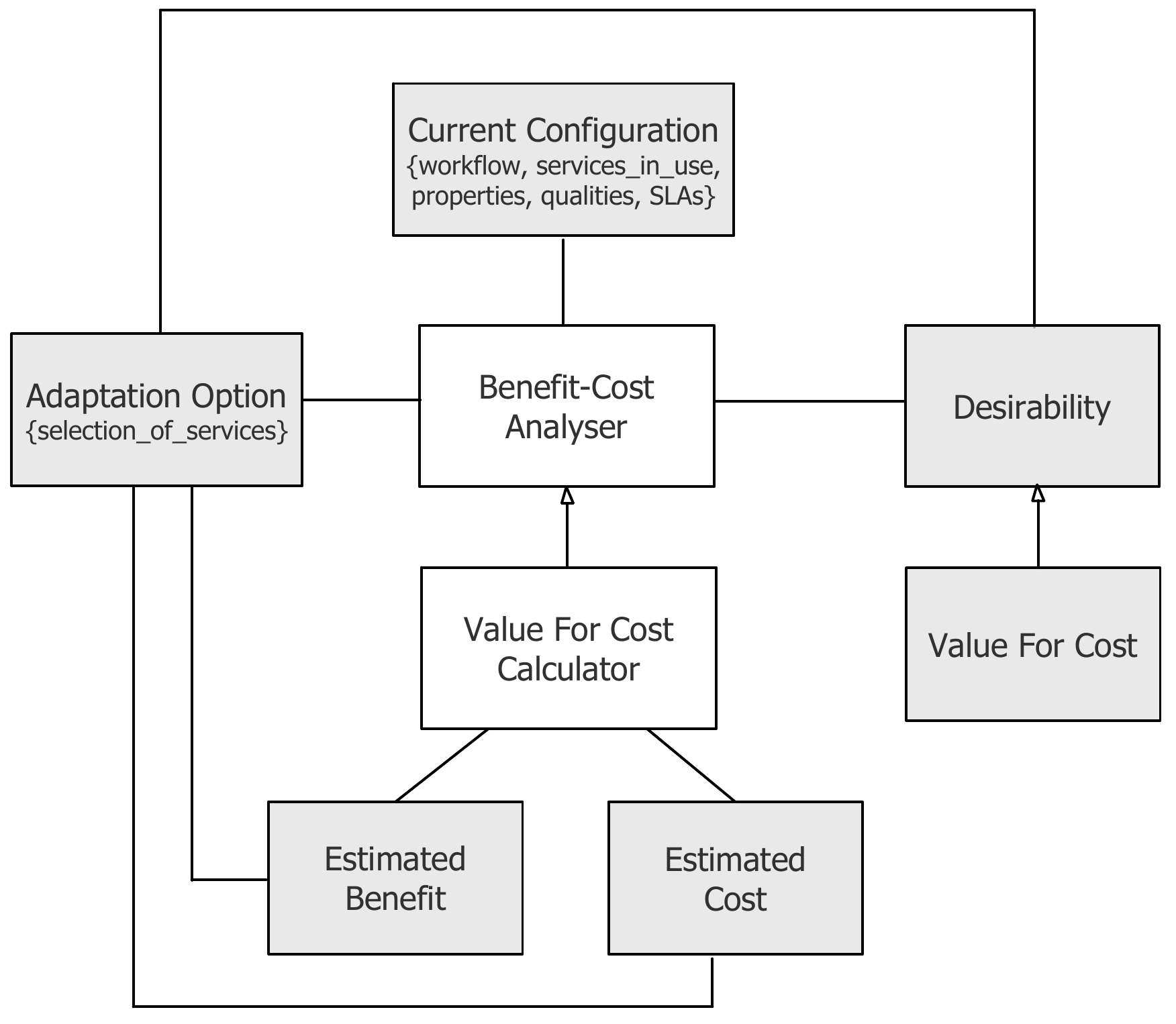}
	\caption{Example instance of Desirability Estimation Model Kind}
	\label{fig:example_MK3}
\end{figure}

\vspace{5pt}\noindent 
\textit{Example.} Figure~\ref{fig:example_MK3} shows the benefit-cost model kind instantiated for the health assistance system. We use value-for-cost (VFC) to express the desirability of adaptation options. The value-for-cost calculator computes VFC for adaptation option $C_i$ as follows: 

\begin{equation}\label{eq5}
VFC_{C_i} = \frac{s_b(EB_{C_i})}{s_c(EC_{C_i})} 
\end{equation}

with $s_b$ a function that scales estimated benefit $EB_{C_i}$ and $s_c$ a function that scales the estimated cost $EC_{C_i}$. In this example, we use trivial scaling functions that return the values of the original estimates, i.e., $s_b(EB_{C_i})= EB_{C_i}$ and $s_c(EC_{C_i})= EB_{C_i}$. 
Applied to the two adaptation options $C_1$ and $C_2$ that we already used to illustrate estimated benefit and estimated cost, we obtain the following values for VFC. 

\begin{equation}\label{eq5}
VFC_{C_1} = \frac{24.5}{4} = 6.13;\mbox{\ \ } VFC_{C_2} = \frac{29.0}{6} = 4.83   
\end{equation}

Although adaptation option $C_2$ has a higher estimated benefit as adaptation option $C_1$, the desirability of adaptation option $C_1$ in terms of value-for-cost is significantly better as for adaptation option $C_2$. The reason is that the estimated cost associated with adapting the current configuration to the new configuration is higher for adaptation option $C_2$ compared to $C_1$. Hence, if an adaptation decision would be made using value-for-cost based on estimated benefit and cost of only adaptation options $C_1$ and $C_2$, adaptation option $C_1$ would be selected for adaptation.

\vspace{5pt}
\noindent\textbf{Risk Estimation Model Kind (MK4).} This model kind describes how the risk of each adaptation option is estimated (see Table~\ref{tab:vp_mk2} left). Risk in general refers to potential effects of uncertainties on system objectives in terms of their likelihoods and consequences (positive or negative or both)~\cite{RiskWebsite}. 
A \textit{risk attribute} represents a particular type of risk that is implied by adapting the managed system with a given adaptation option. Different risk attributes can be associated with applying an adaptation option, such as safety, environment, finances, etc. \textit{Risk metrics} define measures to quantify risk attributes. A \textit{risk estimator} is a mechanism that enables estimating the risk attributes of the adaptation options. A variety of techniques have been established in different domains to estimate risks. Essential to these techniques is that they capture the stakeholders their concerns and intent~\cite{Fischhoff2015}. Example mechanisms include consequence/likelihood matrix, cause-consequence analysis, and decision-tree analysis~\cite{RiskWebsite}. These methods differ in the purpose of the assessment, the information that is requited/available, the importance of the decision, the time available to make a decision, among other criteria. Hence, the choice for a mechanism should be tailored to the context and requirements at hand. The \textit{estimated risk} represents the overall expected risk of applying an adaptation option based on the estimated risk attributes and the combined risk metrics. 
Combining risk \mbox{metrics accounts for the interactions and dependencies between risks.}

To support architects with identifying risk attributes for a problem at hand, we devised a list of high-level risk attributes with associate risk metrics, see Table~\ref{tab:risk_dim_risk_met}. This table is extracted from literature on risks, including~\cite{RiskWebsite,Fischhoff2015,Elmontsri2014,BAYBUTT2015163,Alberts2002,3194710}. 
\begin{table}[htbp]
    \centering
        \caption{Generic risk attributes with possible metrics.}
    \label{tab:risk_dim_risk_met}
    \begin{tabular}{|c|c|}
        \hline
        \textbf{Risk attribute}     & \textbf{Risk metrics} \\ \hline
        Health        &   	Fatalities, aid required  \\ \hline
	    Safety	      &  Fatalities, aid required \\ \hline
	    Security    &  Vulnerability, impact \\ \hline
	    Privacy    &  Data loss, impact \\ \hline
	    Community    &  Outrage, damage \\ \hline
	    Environment    &  Harm, damage \\ \hline
	    Financial	&  Loss, costs \\ \hline
    \end{tabular}
\end{table}

\begin{figure}[h!]
	\centering
	\includegraphics[height=0.4\textwidth]{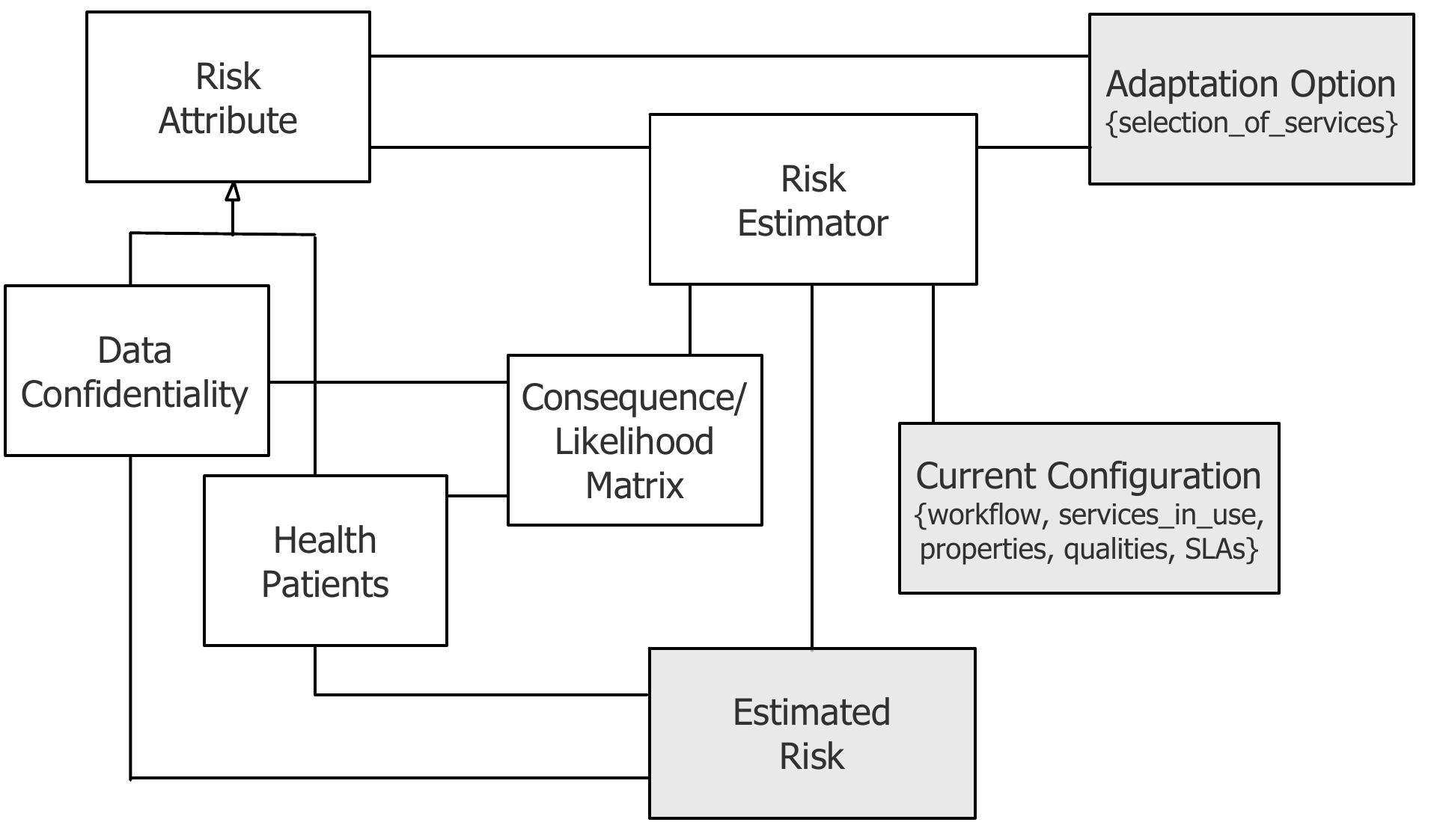}
	\caption{Example instance of Risk Estimation Model Kind}
	\label{fig:example_MKRisk}
\end{figure}

\vspace{5pt}\noindent 
\textit{Example.} Figure~\ref{fig:example_MKRisk} shows an instance of the risk estimation model kind for the health assistance system with two risk attributes: risk on the  confidentially of the data of patients based on exposure of data, and risk on the health of patients based on not 100\% accurate analysis results. We illustrate risk estimation for the health assistance system using a \textit{consequence/likelihood matrix}. 

A consequence/likelihood matrix (or risk matrix) enables specifying an estimated risk according to its consequence and likelihood. The matrix requires customised scales for consequence (Y-axis) and likelihood (X-axis). A common approach is to use discrete scales with three to five points that can be qualitative or quantitative. The scales are determined based on the concerns of the stakeholders and their objectives. The scale for consequences can have positive or negative consequences.

Table~\ref{tab:risk_metric_example} gives an overview of the risk metrics for confidentially of data elicited from the stakeholders.\footnote{The metrics used in this example are inspired by examples provided in the Risk Standard IEC 31010:2019~\cite{RiskWebsite}. Yet, this is just an example for illustration purposes, a wide variety of scales and values may apply depending on the situation at hand.} For instance, a service provider with label gold will store patient data locally. Exposure of data is expected to happen rarely and if it would happen it will have a negligible effect. On the other hand, a service provider with bronze label will share patient data with partners. Consequently, exposure of data is likely and if it would happen it may lead to sensitive data loss. 

\begin{table}[htbp]
    \centering
        \caption{Risk metrics for confidentially of data of the health assistance system.}
    \label{tab:risk_metric_example}
    \begin{tabular}{|c|c|c|c|}
        \hline
        \textbf{SP Label}     & \textbf{Data Policy} & \textbf{Likelihood} & \textbf{Consequence}\\ \hline
        Gold & data stored local & rarely (1/3) & negligible effect (1) \\ \hline
	    Silver	      &  stored with partners & possibly (2/3) & limited impact (2) \\ \hline
	    Bronze   &  shared with partners & likely (3/3) & sensitive data loss (3)  \\ \hline
	    No label    &  not specified & almost certain (4/3) & significant impact (4) \\ \hline
    \end{tabular}
\end{table}

The different options of likelihood and consequence have associated values that allow to determine risk when multiple services are combined (likelihood 1/3 to 4/3, and consequence 1 to 4). We apply the following rules in the example: 

\begin{equation}\label{eqlc}
LC_{C_i} = round(LS_{SPi}^{MAS} + LS_{SPi}^{DS} + LS_{SPi}^{AS})  
\end{equation}

\begin{equation}\label{eq6}
CC_{C_i} = max(CS_{SPi}^{MAS}, CS_{SPi}^{DS}, CS_{SPi}^{AS})  
\end{equation}\vspace{2pt}

The \textit{likelihood} of a combination of services $LC_{C_i}$ of a configuration (adaptation option) is computed by rounding the sum of the likelihood of the individual services. On the other hand, the \textit{consequence} of a combination of services $CC_{C_i}$ is determined by the maximum consequence of any of the services of the configuration. 

\begin{figure}[h!]
	\centering
	\includegraphics[width=0.8\textwidth]{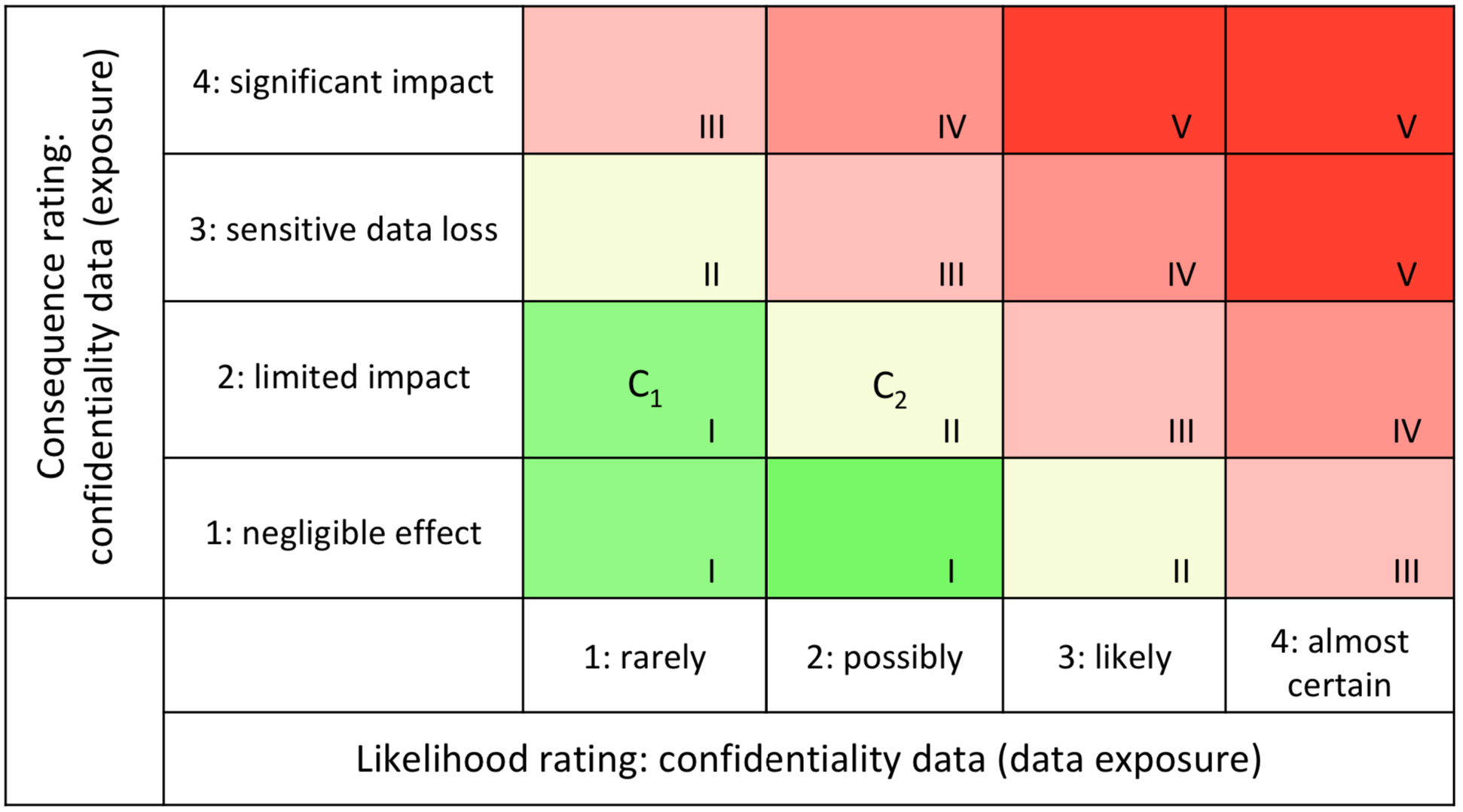}
	\caption{Example of a consequence/likelihood matrix for confidentially of the data.}
	\label{fig:consequence-likelihood-matrix-data}
\end{figure}

Figure~\ref{fig:consequence-likelihood-matrix-data} shows a consequence/likelihood matrix for the risk attribute confidentially of the data. In the example, the scale for consequences for confidentiality of data in terms of ``data exposure'' have a 4-point scale from ``negligible effect'' to ``significant impact.'' Likelihood in terms of ``data exposure'' also have a 4-point scale from ``rarely'' to ``almost certain.'' Each cell of the matrix expresses the estimated risk at one of five possible levels, with level I corresponding to the lowest risk and level V the highest. A risk estimator will map each adaptation option to one cell of the matrix based on the risk metrics determined by the stakeholders (Table~\ref{tab:risk_metric_example}) and the rules defined above. As an example, adaptation option $C_2$ (with a medical analysis service, a drug service, and an alarm service all provided by provider 1) is mapped as follows:

\begin{equation}\label{eqlc}
LC_{C_2} = round(LS_{SP1}^{MAS} + LS_{SP1}^{DS} + LS_{SP1}^{AS}) = round(2/3 + 2/3 + 2/3) = 2   
\end{equation}

\begin{equation}\label{eq6}
CC_{C_2} = max(CS_{SP1}^{MAS}, CS_{SP1}^{DS}, CS_{SP1}^{AS}) = max(2, 2, 2) = 2   
\end{equation}\vspace{2pt}

A value of $LC_{C_2} = 2$ corresponds to likelihood of data exposure ``possibly'' and a value of $CC_{C_2} = 2$ corresponds to an estimated consequence of data exposure ``limited impact.'' As a result, the estimated risk for confidentiality of adaptation option $C_2$ is level II, i.e.,  $ERL_{C_2}^{Data}$ = 2 (see Figure~\ref{fig:consequence-likelihood-matrix-data}).  
Likewise, adaptation option $C_1$ (with a medical analysis service provided by service provider 1, an a drug and alarm service provided by provider 2) can be mapped to likelihood ``rarely'' and consequence ``limited impact,'' with estimated risk for confidentiality 
$C_1$ level I, i.e., $ERL_{C_1}^{Data}$ = 1.
%
%
%

In a similar way, risk metrics and a consequence/likelihood matrix can be defined for risk on the health of patients based on the accuracy of analysis results. 
The overall estimated risk for each adaptation option is then determined by combining the estimated risk per risk attribute. To that end, different approaches can be applied, from basic adding or multiplying elements to providing a magnitude for a risk using a weighting factor to either the consequence or likelihood~\cite{RiskWebsite}. Regardless of the method used, it is important to ensure that the units are consistent and that the impact of a very high risk of one attribute should be treated properly as it may be ``hidden'' by very low risks of the other attributes when combined. To that end, adaptation options with estimated risks above certain thresholds may be ruled out before composing risk attributes.  

Let us determine the overall estimated risk of adaptation options for the health assistance system using a weighted sum as an example. Assume we have a 4 x 4 consequence/likelihood matrix for health of patients with risk levels I to V similar to the consequence/likelihood matrix for data confidentiality. Further, assume that the estimated risk for health of patients of adaptation option $C_1$ is level II, i.e.,  $ERL_{C_1}^{Health}$ = 2, and the risk of adaptation option $C_2$ is $ERL_{C_2}^{Health}$ = 1. With a weight factor $W^{Data} = 0.2$ and $W^{Health} = 0.8$, the overall estimated risk can then be determined as follows:  

\begin{equation}\label{eq5}
ER_{C_1}= ERL_{C_1}^{Data} * W^{Data} + ERL_{C_1}^{Health} * W^{Health} = 1 * 0.2 + 2 * 0.8 = 1.8
\end{equation}

\begin{equation}\label{eq5}
ER_{C_2}= ERL_{C_2}^{Data} * W^{Data} + ERL_{C_2}^{Health} * W^{Health} = 2 * 0.2 + 1 * 0.8 = 1.2
\end{equation}\vspace{2pt}

Hence, in this example, selecting adaptation option $C_2$ for adaptation would result in a lower estimated risk than selecting adaptation option $C_1$. Therefore, $C_2$ would be preferred over $C_1$ if only the estimate risk would matter.
 

\vspace{5pt}
\noindent\textbf{Decision-Making Model Kind (MK5).}
%
The fifth and last model kind describes how adaptation decisions are made (see Table~\ref{tab:vp_mk2} right). A \textit{decision-making mechanism} provides the means to select an adaptation option from the set of available options taking into account the estimated desirability (based on the estimated desirability analysis in terms of estimated benefit and cost), and estimated risk (based on estimated risk analysis). The \textit{selected adaption option} $C_s$ represents the new configuration that is selected to be applied to the managed system in order to adapt it. In general decision-making realises the following abstract function: 

\begin{equation}\label{eq5}
C_s = select(C_c, \{(C_i, ED_{C_i}, ER_{C_i})\}) 
\end{equation}

$\{(C_i, ED_{C_i}, ER_{C_i})\}$ represents the set of triples of all adaptation options $C_i$ together with their associated estimated desirability $ED_{C_i}$ and estimated risk $ER_{C_i}$. 
The $select$ function can be implemented in different ways, from a simple weighted combination of the parameters up to an integrated computation of benefit-cost-risk~\cite{Boardman2011} that may require additional or more detailed data. 

\begin{figure}[h!]
	\centering
	\includegraphics[height=0.5\textwidth]{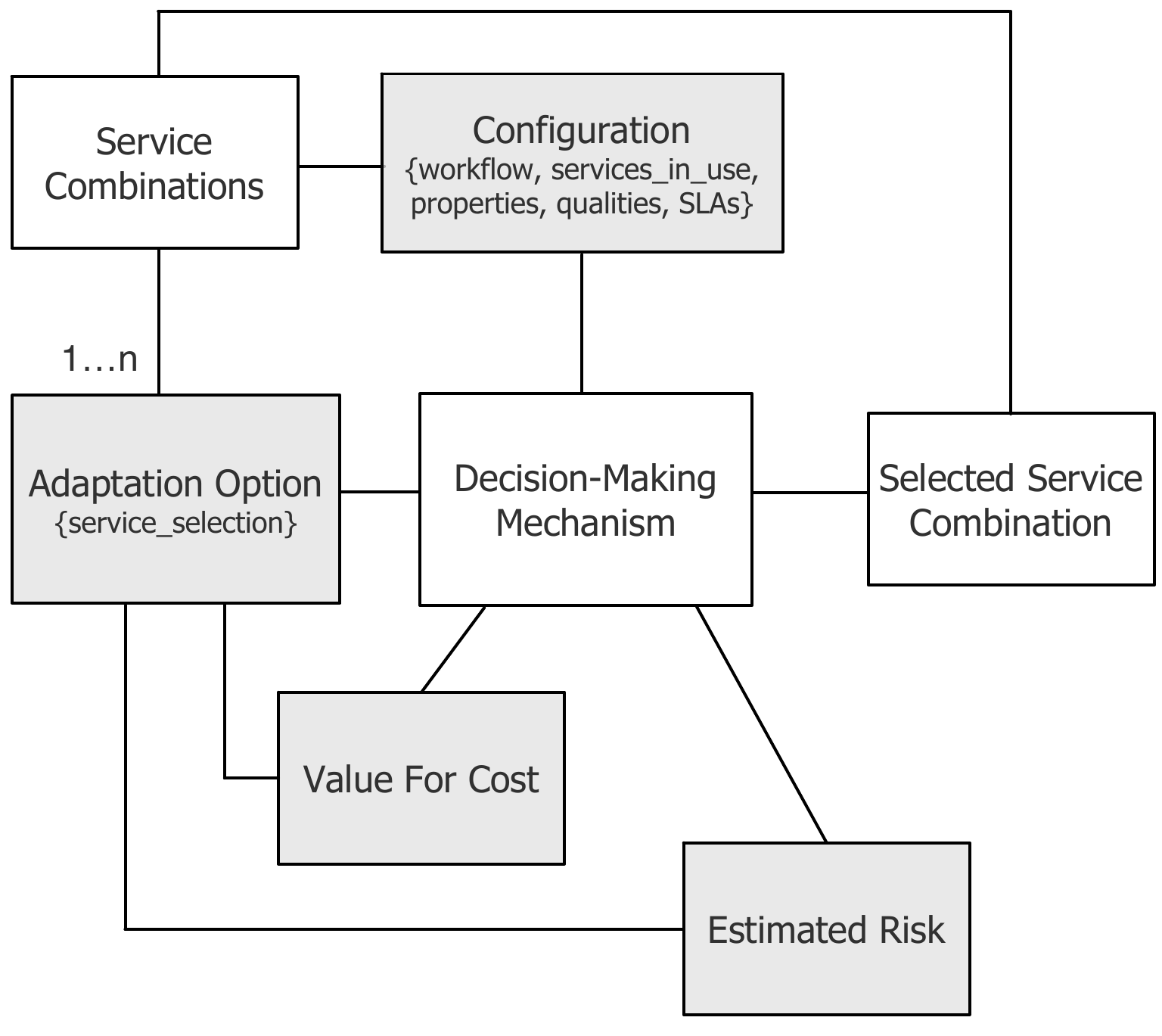}
	\caption{Example instance of Decision-Making Model Kind}
	\label{fig:example_MK4}
\end{figure}

\vspace{5pt}\noindent 
\textit{Example.} Figure~\ref{fig:example_MK4} shows the decision-making model kind instantiated for the health assistance system. We illustrate decision-making using a mechanism that determines the best combination of services from the possible service configurations based on a weighted combination of estimated value-for-cost and risk. 

The $select$ function is implemented as follows: 

\begin{equation}\label{eq5}
EBCR_{C_s} = \max\limits_{}\sum_{i=1}^{n}  (ED_{C_i} * W_{VFC} -  ER_{C_i} * W_{R})
\end{equation}

The adaptation option $C_s$ that is selected for adaption is the option that maximises the estimated benefit-cost-risk $EBCR_{C_s}$. $EBCR_{C_s}$ is computed as the  difference between the weighted estimated desirability $ED_{C_i} * W_{VFC}$ and the weighted estimated risk $ER_{C_i} * W_{R}$. 

The weights $W_{VFC}$ and $W_{R}$ determine the importance stakeholders put in the estimated desirability of the selected adaptation option, i.e., its value-for-cost in the health assistance system, and the estimated risk. If we assume equal weights, i.e., $W_{VFC} = 0.5$ and $W_{R} = 0.5$, and we use the values for estimated desirability and risk for $C_1$ and $C_2$ from the previous examples, the decision will be made as follows: 

\begin{equation}\label{eq5}
EBCR_{C_1} = ED_{C_1} * W_{VFC} -  ER_{C_1} * W_{R} = 6.13 * 0.5 - 1.8 * 0.5 = 2.17\vspace{-5pt}
\end{equation}

\begin{equation}\label{eq5}
EBCR_{C_2} = ED_{C_2} * W_{VFC} -  ER_{C_2} * W_{R} = 4.83 * 0.5 - 1.2 * 0.5 = 1.81\vspace{5pt}
\end{equation}

If $C_1$ and $C_2$ would be the only two adaptation options to select from, the decision-making mechanism would select adaptation option $C_1$
over $C_2$. However, if the stakeholders would prefer different  weights for estimated desirability and risk, the outcome may be different. We elaborate on this in the next section. 


\subsection{Viewpoint Analysis}

The viewpoint defines two types of analysis presented in Table~
\ref{tab:analysis}: \textit{benefit-cost tradeoff analysis} and \textit{desirability-risk tradeoff analysis}. 

\begin{table}[!h]
\renewcommand{\arraystretch}{1.2}
\caption{Viewpoint -- Analysis}\vspace{-5pt}
\label{tab:analysis}
\centering
\begin{tabular}{ p{13.4cm} }
\toprule
\textbf{Analyses:}\\
\textit{A1 - Benefit-Cost Tradeoff Analysis (using MK3):} Assesses the effects of different weights for benefit and cost on the overall estimated desirability of adaptation options of a given configuration with a given benefit-cost analysis mechanism.\\

\textit{A2 - Desirability-Risk Tradeoff Analysis (using MK5):} Assesses 
the effects of different weights for desirability and risk on the selection of adaptation options of a given current configuration with a given decision-making mechanism. \\
\bottomrule
\end{tabular}
\end{table}

\vspace{5pt}
\noindent\textbf{Benefit-Cost Tradeoff Analysis.} 
This analysis is applied to a selection of relevant adaptation scenarios, each comprising a current configuration with a selection of adaptation options. The analysis then assesses the effects of assigning different weights to the estimated benefit and estimated cost on the desirability of the adaptation options. The results can then be checked with domain knowledge obtained from stakeholders, historical information, field tests, or any other relevant data sources. The analysis results help balancing the tradeoffs between estimated benefit and estimated cost when designing the benefit-cost analyser. This analysis is usually performed at design time, but may also be useful in the context of a system evolution; for instance when new goals or risks are identified that need to be incorporated into the self-adaptive system. 

\vspace{5pt}
\noindent\textit{Example.}
We illustrate benefit-risk tradeoff analysis for the simple scenario of the health assistance system that we used to illustrate the benefit-cost model kind. In that example we used trivial scaling functions that return the values of the original estimates for benefit and cost, i.e., $s_b(EB_{C_i})= EB_{C_i}$ and $s_c(EC_{C_i})= EB_{C_i}$ and we applied that to determine estimated VFC of two adaptation options ($C_1$ and $C_2$).

We look now how a parametric scaling function for benefit gives preference to adaptation options with high benefit as follows: 

\begin{equation}\label{eq5}
VFC_{C_i} = \frac{T + (EB_{C_i} - T) * x}{EC_{C_i}} 
\end{equation}

\begin{figure}[h!]
	\centering
	\includegraphics[height=0.4\textwidth]{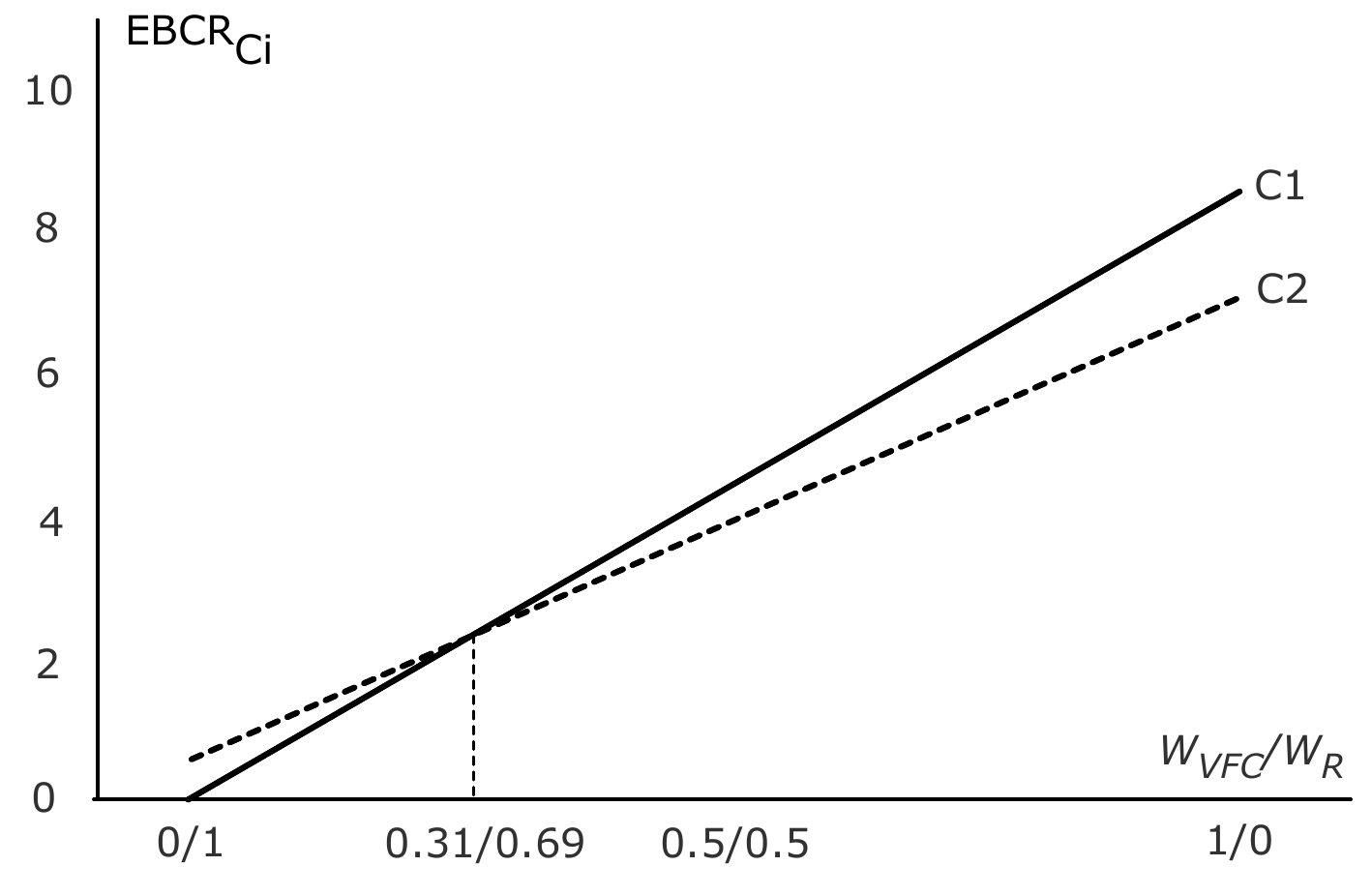}
	\caption{Changing decision based on weights for estimated desirability (VFC) and risk.}
	\label{fig:example_decision_weigths}
\end{figure}

The scaling function of estimated benefit is determined based on $T$, a threshold for estimated benefit, and a multiplier $x$. The scaling function of estimated cost remains trivial, returning the values of the original estimates for cost. The overall estimated benefit of an adaptation option is determined by making the sum of the threshold and the fraction of the estimate above the threshold multiplied with a factor $x$. 
Figure~\ref{fig:example_decision_weigths} shows how estimated value-for-cost ($VFC_{C_i}$) is determined for two adaptation options $C_1$ and $C_2$. 

\begin{figure}[h!]
	\centering
	\includegraphics[height=0.4\textwidth]{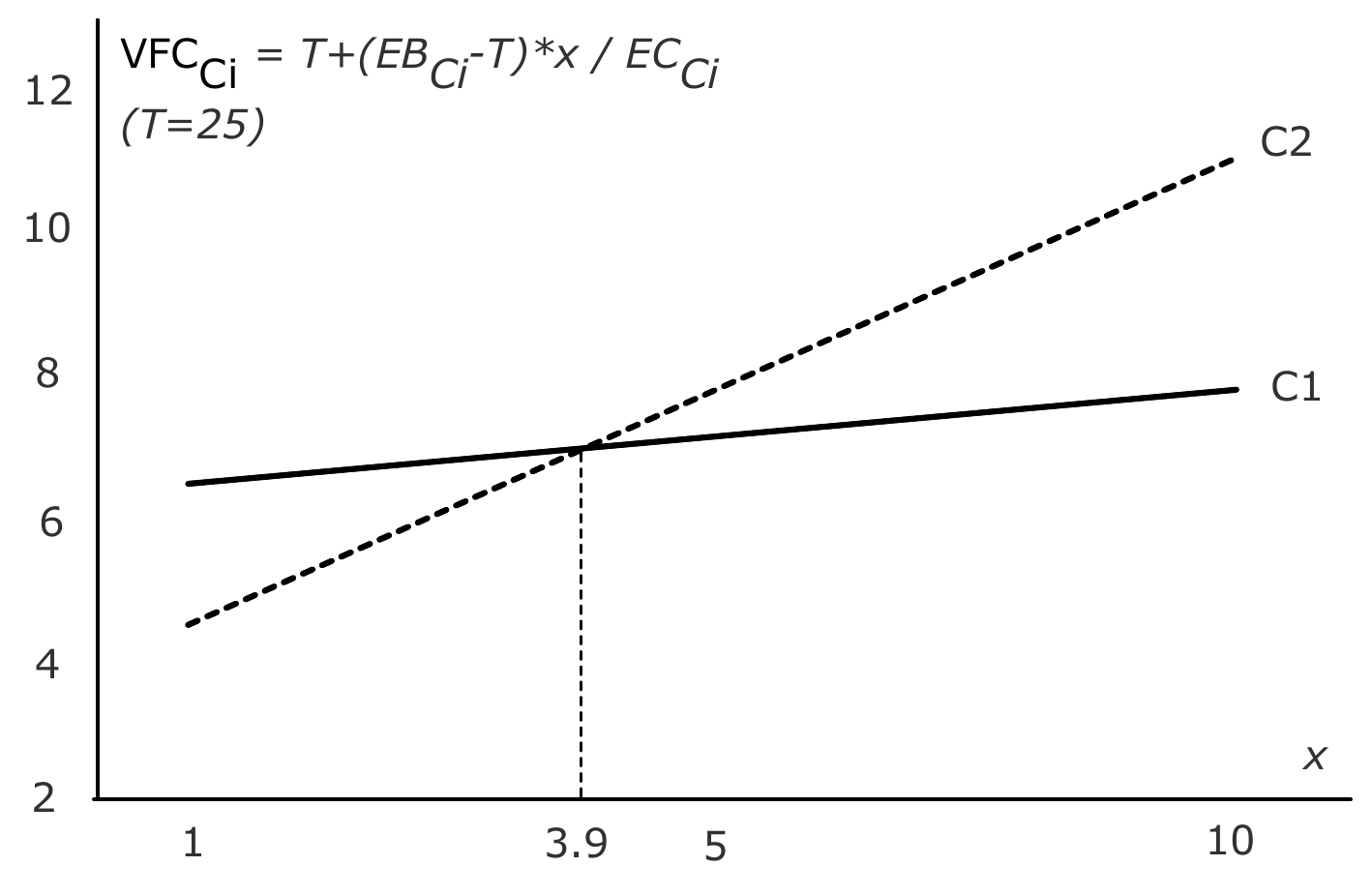}
	\caption{Change of desirability (VFC) based on weights for estimated benefit and cost.}
	\label{fig:example_VFC}
\end{figure}

In this particular setting, threshold $T$ is set to 25 and $x$ is changed in a range from 1 to 10. Note that the setting with $x = 1$ corresponds to the original setting we used to illustrate the desirability model kind ($VFC_{C_1} = 6.13$ and $VFC_{C_2} = 4.83$). 
As we can see, for values of $x$ equal or smaller than $3.9$ the value-for-cost of adaptation option $C_1$ would be preferred over that of $C_2$. For the complementary range of values the opposite choice would be preferred. This means that above $x = 3.9$ the higher contribution of estimated benefit of $C_2$ above the threshold (namely 29.0 compared to 24.5 for $C_1$), would compensate the higher cost (6 for $C_2$ versus 4 for $C_1$). The final choice for the threshold and multiplication factor $x$ is something the stakeholders need to make. 

\vspace{5pt}
\noindent\textbf{Desirability-Risk Tradeoff Analysis.}
This analysis is applied to a selection of relevant adaptation scenarios, each comprising a configuration and a set of adaptation options. The analysis assesses the effects of assigning different weights to estimated desirability and risk on the selection of adaptation options by the decision-making mechanism. The results can then be compared with domain knowledge. The analysis results help balancing estimated desirability (based on estimated benefit and cost) and estimated risk in the decision-making process. The results will help determining the knowledge required by the decision-making mechanism. Desirability-risk tradeoff analysis is usually performed at design time, but it can also be useful during system evolution. 

\vspace{5pt}
\noindent\textit{Example.}
We illustrate the desirability-risk tradeoff analysis for the simple scenario of the health assistance system that we used to illustrate the decision making model kind. In that example we assumed equal weights, i.e., $W_{VFC} = 0.5$ and $W_{R} = 0.5$ to select one of two possible adaptation options ($C_1$ and $C_2$). 

We look now how a change of the weights have an effect on the decision-making. Figure~\ref{fig:example_decision_weigths} shows how estimated desirability-risk of the two adaptation options ($EBCR_{C_1}$ and  $EBCR_{C_2}$) change with different weights. 
For values of $W_{VFC}$ equal or smaller than threshold X (e.g., 0.31 and hence, values of $W_{R}$ approximately equal or larger than 0.69) adaptation option $C_2$ would be preferred over $C_1$.This means, the stakeholders would give much more attention to risk as to desirability in terms of value-for-cost. For the complementary range of values for the weights the opposite choice would be made. The final choice for the weights is something the stakeholders need to make. 

\section{Conclusion}
\label{sec:concl}

In this paper, we presented a specification of an architecture viewpoint for benefit-cost-risk-aware decision-making in self-adaptive systems, aligned with ISO/IEC/IEEE 42010. The viewpoint is intended to address the concerns of architects, system owners, users and other stakeholders of  self-adaptive systems. The novelty of the viewpoint lays in the combination of estimated benefit, cost, and risk as first-class citizens in the decision-making process to select configurations to adapt the system. We devised the viewpoint to be flexible in the mechanisms used to make different decisions. The viewpoint is also intended to be compatible with other architectural approaches that are required to design other aspects of a self-adaptive system, such as monitoring of the system and its environment, the generation of plans, and the execution of these plans to enact adaptations.


%
%
%
\bibliographystyle{ACM-Reference-Format}
\bibliographystyle{splncs04}
\bibliography{references}
%


%
%
%
%
\newpage 

\section*{Appendix A: Terminology}
\label{sec:terminology}

\textit{Current configuration} represents the actual aspects of the managed system and the environment that are relevant to adaptation. This includes the current configuration of components of the managed system, the current values of the properties of interest, and the current values of uncertainties that are relevant to adaptation. In the service-based system example, the current configuration consists of the current set of services used in the workflow, the actual values associated with the different paths in the workflow, and the current values of properties such as reliability, and service level agreements.  

\textit{Adaptation options} are the possible configurations that can be reached by adapting the current configuration of the system. In the service-based system, the adaptation options are determined by the combination of all concrete services provided by the service providers that can be composed to in the workflow. 

\textit{Adaptation goals} represent the quality requirements that need to be achieved by the managing system. In the service-based system example, the adaptation goals are failure rate and resources usage service invocations. 

\textit{Benefit attributes} of an adaptation option are the estimated values for each property that corresponds with an adaptation goal. These properties are usually quality properties of the system. In the service-based system example, the quality attributes are the estimated failure rate and estimated resource usage of an adaptation option that is expected to be achieved when the system is adapted according to that adaptation option. 

\textit{Benefit estimator} is a mechanism that is used to estimate the benefit of  adaptation options. In the service-based system example, we may use an estimator that evaluates the utility of adaptation options by simulating the workflow of different adaptation options using the actual values associated with the different paths in the workflow.  

\textit{Estimated benefit} represents the estimated overall benefit of an adaptation option based on the estimated benefit attributes and the associated adaptation goals. In the service-based system example, the benefit may be represented as an utility that is determined as a sum of the weighted values of the estimated quality attributes: failure rate and resource usage. 

\textit{Cost dimensions} represent the different costs that are implied by adaptation of the managed system. In the service-based system example, we may consider the overhead that is implied by testing new concrete services that are included in the service workflow. 

\textit{Cost metrics} define measures to quantify the cost dimensions. For instance, testing new services requires resources that may depend on the service level agreements made with the service providers. 

\textit{Cost estimator} is a mechanism that is used to estimate the cost attributes of the adaptation options. In the service-based system example, the cost estimator may determine the expected resources required to test new services when selecting different adaptation options. 

\textit{Estimated cost} represents the estimated overall cost of an adaptation option based on the estimates for the different cost attributes and their relevance. In the service-based system, the estimated cost corresponds to the resources that are expected to be required to test new services of the selected adaptation option. 

\textit{Benefit-cost analyser} determines the desirability of an adaptation option. Computing desirability requires that the estimated benefit and estimated cost are expressed in a common metric and are scaled  to be comparable. For the service-based system the desirability of adaptation options can be computed as value-for-cost that determines desirability as the scaled estimated benefit over the scaled estimated cost. 

\textit{Estimated desirability} expresses the degree that stakeholders prefer one adaptation option over the other options by comparing overall estimated benefit with overall estimated cost. For the service-based system the value-for-cost can be used to represent the estimated desirability of adaptation options. 

\textit{Risk attributes} represent the different types of risk that are implied by adapting a managed system with a given adaptation option. Risk attributes can related to safety, environment, finances, etc. In the service-based system,
the confidentiality of data and the health of patients are two important risk attributes. 

\textit{Risk metrics} define measures to quantify risk attributes. A consequence/likelihood matrix is an example to represent risk metrics. Such a matrix enables specifying an estimated risk according to its consequence and likelihood based on stakeholder input. In the service-based system example, the likelihood of data exposure (for the risk of confidentiality of data) may range over a 4-point scale from rarely to almost certain, while the consequences can range from negligible effect to significant impact.  

\textit{Risk estimator} is a mechanism that is used to estimate the risk attributes of the adaptation options. In the service-based system example, the estimated risk may be determined based on the service level agreements with the providers of the selected services of the adaptation options, based on the likelihood/consequence matrices of the different risk attributes.  

\textit{Estimated risk} of an adaptation option represents the estimated overall risk of an adaptation option based on the estimates for the different risk attributes and the weights associated with them. In the service-based system example, the risk may be represented as a weighted sum of the two risk attributes related to data exposure and health of patients.  

\textit{Decision-making mechanism} is a mechanism that selects an adaptation option from the set of available options taking into account the estimated desirability and estimated risk of all available adaptation options.  In the service-based system example, the decision-making mechanism may select the adaptation option that maximises the difference between weighted desirability and risk. 

\textit{Selected adaption option} represents the new configuration that is selected for adaptation and will be applied to the managed system. In the service-based system example, the selected adaptation option comprises the set of service that the workflow needs to invoke; some of the current services may remain in use, others may need to be replaced by new selected services.

\end{document}